\documentclass[nofootinbib,notitlepage]{revtex4-1}
\usepackage{verbatim}
\usepackage{graphicx} 
\usepackage{dcolumn}  
\usepackage{bm}       
\usepackage{amssymb}

\usepackage[english]{babel}
\usepackage[normalem]{ulem}
\usepackage{color}
\usepackage[hypcap]{caption}
\usepackage{amsmath}
\usepackage[mathscr]{eucal}
\DeclareMathAlphabet{\mathpzc}{OT1}{pzc}{m}{it}

\usepackage[]{hyperref}

\hypersetup{
  pdftitle={Your title here},
  pdfauthor={Your name here},
  pdfsubject={Your subject here},
  pdfkeywords={keyword1, keyword2},
  bookmarksnumbered=true,     
  bookmarksopen=true,         
  bookmarksopenlevel=1,       
  colorlinks=true,            
  pdfstartview=Fit,           
  pdfpagemode=UseOutlines,
  pdfpagelayout=TwoPageRight
}

\usepackage{natbib}

\usepackage{oubraces} 
\usepackage{mathtools} 

\usepackage{indentfirst}
\usepackage{setspace}

\usepackage{ifthen}

  \renewcommand*\And
    {\mathop{}\!\wedge}
  \newcommand*\Or
    {\mathop{}\!\vee}
  \newcommand*\If
    {\mathop{}\!\Rightarrow}
  \newcommand*\fI
    {\mathop{}\!\Leftarrow}
  \newcommand*\Iff
    {\mathop{}\!\Leftrightarrow}
  \newcommand*\Xor
    {\mathop{}\!\veebar}

  \newcommand{\fa}[2]
    {\ensuremath{\forall_{#1}\left(#2\right)}}
  \newcommand{\ex}[2]
    {\ensuremath{\exists_{#1}\left(#2\right)}}
  \newcommand{\exU}[2]
    {\ensuremath{\exists\! !_{#1}\left(#2\right)}}
  \newcommand{\exC}[3]
    {\ensuremath{\exists^#1_{#2}\left(#3\right)}}
  \newcommand{\st}[2]
    {\ensuremath{#1\backepsilon_{#2}}}
  \newcommand{\subst}[3]
    {\ensuremath{\left#3\right\vert_{#1 := #2}}}
  \newcommand{\delim}[3]    
  {
    \ifthenelse{\equal{#1}{}}
    {
      \ifthenelse{\equal{#2}{}}
      {
        #3
      }
      {
        \left. #3 \right #2
      }
    }
    {
      \ifthenelse{\equal{#2}{}}
      {
        \left #1 #3 \right.
      }
      {
        \left #1 #3 \right #2
      }
    }
  }

  \newcommand*{\defeq}{\stackrel{\text{def}}{=}}
  \newcommand{\esteq}
    {\mathrel{\hat{=}}}
    
  \newcommand{\dom}[1]
    {\mathrm{D}_{#1}}
  \newcommand{\im}[1]
    {\mathrm{Im}_{#1}}
  \newcommand{\cdom}[1]
    {\mathrm{CD}_{#1}}
  \newcommand{\Conj}[3]
  {
    \ifthenelse{\equal{#2}{}}
    {
      \ifthenelse{\equal{#3}{}}
      {
        \mathbb{#1}
      }{
        \Conj{#1}{}{}^{#3}
      }
    }{
      \ifthenelse{\equal{#3}{}}
      {
        \Conj{#1}{}{}_{#2}
      }{
        \Conj{#1}{}{}_{#2}^{#3}
      }
    }
  }
  \newcommand{\set}[3]
    {\ensuremath{\mathbb{#1}_{#2}^{#3}}}

  \newcommand{\sgn}[1]
    {\mathrm{sgn}\left(#1\right)}
  \providecommand{\sin}{} \renewcommand{\sin}{\mathrm{sen}}

  \newcommand{\Ind}[2] {
    \ifthenelse{\equal{#2}{}}
    {
      \chi_{#1}
    }
    {
      \chi_{#1}\left(#2\right)
    }
  }

  \newenvironment{xsmallmatrix}
    {\renewcommand\thickspace{\kern0em}\smallmatrix}
    {\endsmallmatrix}
  \newcommand{\Matrix}[8]    
  {
    \ifthenelse{\equal{#1}{}}{
      \begingroup
      \ensuremath{
        \ifthenelse{\equal{#4}{}}{\arraycolsep=5pt}{\arraycolsep=#4} 
        \ifthenelse{\equal{#5}{}}{\def\arraystretch{2}}{\def\arraystretch{#5}} 
        \left #2 
        \begin{array}{*{#6}{#7}} #8 \end{array}
        \right #3
      }
      \endgroup
    }{
      \begingroup
      \scalebox{#1}{
        \ensuremath{
          \ifthenelse{\equal{#4}{}}{\arraycolsep=5pt}{\arraycolsep=#4} 
          \ifthenelse{\equal{#5}{}}{\def\arraystretch{2}}{\def\arraystretch{#5}} 
          \left #2 
          \begin{array}{*{#6}{#7}} #8 \end{array}
          \right #3
        }
      }
      \endgroup
    }
  }
  
  \newcommand{\Prb}[6]
  {
    \ifthenelse{\equal{#1}{}}
    {
      \ensuremath{\mathpzc{P}}
    }{}
    \ifthenelse{\equal{#1}{d}}
    {
      \ensuremath{\mathpzc{p}}
    }{}
    \ifthenelse{\equal{#1}{<}}
    {
      \ensuremath{\mathscr{P}}
    }{}
    \ifthenelse{\equal{#1}{_}}
    {
      \Prb{}{}{}{}{}{}_{#2}
    }{}
    \ifthenelse{\equal{#1}{_(}}
    {
      \Prb{}{}{}{}{}{}_{#2}\left(#4\right)
    }{}
    \ifthenelse{\equal{#1}{_(=}}
    {
      \Prb{}{}{}{}{}{}_{#2}\left(#4\right)
    }{}
    \ifthenelse{\equal{#1}{_(;}}
    {
      \Prb{}{}{}{}{}{}_{#2}\left(#4;#6\right)
    }{}
    \ifthenelse{\equal{#1}{_(=;}}
    {
      \Prb{}{}{}{}{}{}_{#2}\left(#4;#6\right)
    }{}
    \ifthenelse{\equal{#1}{(}}
    {
      \Prb{}{}{}{}{}{}\left(#4\right)
    }{}
    \ifthenelse{\equal{#1}{(;}}
    {
      \Prb{}{}{}{}{}{}\left(#4;#6\right)
    }{}
    \ifthenelse{\equal{#1}{d_}}
    {
      \Prb{d}{}{}{}{}{}_{#2}
    }{}
    \ifthenelse{\equal{#1}{d_(}}
    {
      \Prb{d}{}{}{}{}{}_{#2}\left(#4\right)
    }{}
    \ifthenelse{\equal{#1}{d_(;}}
    {
      \Prb{d}{}{}{}{}{}_{#2}\left(#4;#6\right)
    }{}
    \ifthenelse{\equal{#1}{d(}}
    {
      \Prb{d}{}{}{}{}{}\left(#4\right)
    }{}
    \ifthenelse{\equal{#1}{d(;}}
    {
      \Prb{d}{}{}{}{}{}\left(#4;#6\right)
    }{}
    \ifthenelse{\equal{#1}{;_}}
    {
      \Prb{}{}{}{}{}{}_{\left.#2 \vphantom{#3}\right| #3}
    }{}
    \ifthenelse{\equal{#1}{;_(;}}
    {
      \Prb{;_}{#2}{#3}{}{}{}\left(#4 \vphantom{#5}\right| \left. #5 \vphantom{#4}\right)
    }{}
    \ifthenelse{\equal{#1}{;_(;=}}
    {
      \Prb{;_}{#2}{#3}{}{}{}\left(#4;#5\right)
    }{}
    \ifthenelse{\equal{#1}{;_(=;}}
    {
      \Prb{;_}{#2}{#3}{}{}{}\left(#4;#5\right)
    }{}
    \ifthenelse{\equal{#1}{;_(=;=}}
    {
      \Prb{;_}{#2}{#3}{}{}{}\left(#4;#5\right)
    }{}
    \ifthenelse{\equal{#1}{;_(;;}}
    {
      \Prb{;_}{#2}{#3}{}{}{}\left(#4 \vphantom{#5 #6}\right| \left. #5 \vphantom{#4};#6\right)
    }{}
    \ifthenelse{\equal{#1}{;_(;=;}}
    {
      \Prb{;_}{#2}{#3}{}{}{}\left(#4;#5;#6\right)
    }{}
    \ifthenelse{\equal{#1}{;_(=;;}}
    {
      \Prb{;_}{#2}{#3}{}{}{}\left(#4;#5;#6\right)
    }{}
    \ifthenelse{\equal{#1}{;_(=;=;}}
    {
      \Prb{;_}{#2}{#3}{}{}{}\left(#4;#5;#6\right)
    }{}
    \ifthenelse{\equal{#1}{;(;}}
    {
      \Prb{}{}{}{}{}{}\left(#4 \vphantom{#5}\right| \left. #5 \vphantom{#4}\right)
    }{}
    \ifthenelse{\equal{#1}{;(;;}}
    {
      \Prb{}{}{}{}{}{}\left(#4;#5;#6\right)
    }{}
    \ifthenelse{\equal{#1}{d;_}}
    {
      \Prb{d}{}{}{}{}{}_{#2;#3}
    }{}
    \ifthenelse{\equal{#1}{d;_(;}}
    {
      \Prb{d;_}{#2}{#3}{}{}{}\left(#4;#5\right)
    }{}
    \ifthenelse{\equal{#1}{d;_(;=}}
    {
      \Prb{d;_}{#2}{#3}{}{}{}\left(#4;#5\right)
    }{}
    \ifthenelse{\equal{#1}{d;_(;;}}
    {
      \Prb{d;_}{#2}{#3}{}{}{}\left(#4;#5;#6\right)
    }{}
    \ifthenelse{\equal{#1}{d;_(;=;}}
    {
      \Prb{d;_}{#2}{#3}{}{}{}\left(#4;#5;#6\right)
    }{}
    \ifthenelse{\equal{#1}{d;(;}}
    {
      \Prb{d}{}{}{}{}{}\left(#4;#5\right)
    }{}
    \ifthenelse{\equal{#1}{d;(;;}}
    {
      \Prb{d}{}{}{}{}{}\left(#4;#5;#6\right)
    }{}
    \ifthenelse{\equal{#1}{<_}}
    {
      \Prb{<}{}{}{}{}{}_{#2}
    }{}
    \ifthenelse{\equal{#1}{<_(}}
    {
      \Prb{<_}{#2}{}{}{}{}\left(#4\right)
    }{}
    \ifthenelse{\equal{#1}{<_(;}}
    {
      \Prb{<_}{#2}{}{}{}{}\left(#4;#6\right)
    }{}
    \ifthenelse{\equal{#1}{<(}}
    {
      \Prb{<}{}{}{}{}{}\left(#4\right)
    }{}
    \ifthenelse{\equal{#1}{<(;}}
    {
      \Prb{<}{}{}{}{}{}\left(#4;#6\right)
    }{}
    \ifthenelse{\equal{#1}{<;_}}
    {
      \Prb{<}{}{}{}{}{}_{#2;#3}
    }{}
    \ifthenelse{\equal{#1}{<;_(;}}
    {
      \Prb{<;_}{#2}{#3}{}{}{}\left(#4;#5\right)
    }{}
    \ifthenelse{\equal{#1}{<;_(;=}}
    {
      \Prb{<;_}{#2}{#3}{}{}{}\left(#4;#5\right)
    }{}
    \ifthenelse{\equal{#1}{<;_(;;}}
    {
      \Prb{<;_}{#2}{#3}{}{}{}\left(#4;#5;#6\right)
    }{}
    \ifthenelse{\equal{#1}{<;_(;=;}}
    {
      \Prb{<;_}{#2}{#3}{}{}{}\left(#4;#5;#6\right)
    }{}
    \ifthenelse{\equal{#1}{<;(;}}
    {
      \Prb{<}{}{}{}{}{}\left(#4;#5\right)
    }{}
    \ifthenelse{\equal{#1}{<;(;;}}
    {
      \Prb{<}{}{}{}{}{}\left(#4;#5;#6\right)
    }{}
  }
  \newcommand{\PrbEsp}[6]
  {
    \ifthenelse{\equal{#1}{}}
    {
      \ensuremath{\mathpzc{E}}
    }{}
    \ifthenelse{\equal{#1}{(}}
    {
      \PrbEsp{}{}{}{}{}{}\left(#4\right)
    }{}
    \ifthenelse{\equal{#1}{(;}}
    {
      \PrbEsp{}{}{}{}{}{}\left(#4;#6\right)
    }{}
    \ifthenelse{\equal{#1}{_}}
    {
      \PrbEsp{}{}{}{}{}{}_{#2}
    }{}
    \ifthenelse{\equal{#1}{_(}}
    {
      \PrbEsp{}{}{}{}{}{}_{#2}\left(#4\right)
    }{}
    \ifthenelse{\equal{#1}{_(;}}
    {
      \PrbEsp{}{}{}{}{}{}_{#2}\left(#4;#6\right)
    }{}
    \ifthenelse{\equal{#1}{;_}}
    {
      \PrbEsp{}{}{}{}{}{}_{\left.#2 \vphantom{#3}\right| #3}
    }{}
    \ifthenelse{\equal{#1}{;_(;}}
    {
      \PrbEsp{;_}{#2}{#3}{}{}{}\left(#4 \vphantom{#5}\right| \left. #5 \vphantom{#4}\right)
    }{}
    \ifthenelse{\equal{#1}{;_(;=}}
    {
      \PrbEsp{;_}{#2}{#3}{}{}{}\left(#4;#5\right)
    }{}
    \ifthenelse{\equal{#1}{;_(;;}}
    {
      \PrbEsp{;_}{#2}{#3}{}{}{}\left(#4 \vphantom{#5 #6}\right| \left. #5 \vphantom{#4};#6\right)
    }{}
    \ifthenelse{\equal{#1}{;_(;=;}}
    {
      \PrbEsp{;_}{#2}{#3}{}{}{}\left(#4;#5;#6\right)
    }{}
    \ifthenelse{\equal{#1}{;(;}}
    {
      \PrbEsp{}{}{}{}{}{}\left(#4 \vphantom{#5}\right| \left. #5 \vphantom{#4}\right)
    }{}
    \ifthenelse{\equal{#1}{;(;;}}
    {
      \PrbEsp{}{}{}{}{}{}\left(#4 \vphantom{#5 #6}\right| \left. #5 \vphantom{#4};#6\right)
    }{}
  }
  \newcommand{\PrbVar}[6]
  {
    \ifthenelse{\equal{#1}{}}
    {
      \ensuremath{\mathpzc{V}}
    }{}
    \ifthenelse{\equal{#1}{(}}
    {
      \PrbVar{}{}{}{}{}{}\left(#4\right)
    }{}
    \ifthenelse{\equal{#1}{(;}}
    {
      \PrbVar{}{}{}{}{}{}\left(#4;#6\right)
    }{}
    \ifthenelse{\equal{#1}{_}}
    {
      \PrbVar{}{}{}{}{}{}_{#2}
    }{}
    \ifthenelse{\equal{#1}{_(}}
    {
      \PrbVar{}{}{}{}{}{}_{#2}\left(#4\right)
    }{}
    \ifthenelse{\equal{#1}{_(;}}
    {
      \PrbVar{}{}{}{}{}{}_{#2}\left(#4;#6\right)
    }{}
    \ifthenelse{\equal{#1}{;_}}
    {
      \PrbVar{}{}{}{}{}{}_{#2;#3}
    }{}
    \ifthenelse{\equal{#1}{;_(;}}
    {
      \PrbVar{;_}{#2}{#3}{}{}{}\left(#4;#5\right)
    }{}
    \ifthenelse{\equal{#1}{;_(;=}}
    {
      \PrbVar{;_}{#2}{#3}{}{}{}\left(#4;#5\right)
    }{}
    \ifthenelse{\equal{#1}{;_(;;}}
    {
      \PrbVar{;_}{#2}{#3}{}{}{}\left(#4;#5;#6\right)
    }{}
    \ifthenelse{\equal{#1}{;_(;=;}}
    {
      \PrbVar{;_}{#2}{#3}{}{}{}\left(#4;#5;#6\right)
    }{}
    \ifthenelse{\equal{#1}{;(;}}
    {
      \PrbVar{}{}{}{}{}{}\left(#4;#5\right)
    }{}
    \ifthenelse{\equal{#1}{;(;;}}
    {
      \PrbVar{}{}{}{}{}{}\left(#4;#5;#6\right)
    }{}
  }
  \newcommand{\PrbCov}[6]
  {
    \ifthenelse{\equal{#1}{}}
    {
      \ensuremath{\mathrm{Cov}}
    }{}
    \ifthenelse{\equal{#1}{(}}
    {
      \PrbCov{}{}{}{}{}{}\left(#4\right)
    }{}
    \ifthenelse{\equal{#1}{(;}}
    {
      \PrbCov{}{}{}{}{}{}\left(#4;#6\right)
    }{}
    \ifthenelse{\equal{#1}{_}}
    {
      \PrbCov{}{}{}{}{}{}_{#2}
    }{}
    \ifthenelse{\equal{#1}{_(}}
    {
      \PrbCov{}{}{}{}{}{}_{#2}\left(#4\right)
    }{}
    \ifthenelse{\equal{#1}{_(;}}
    {
      \PrbCov{}{}{}{}{}{}_{#2}\left(#4;#6\right)
    }{}
    \ifthenelse{\equal{#1}{;_}}
    {
      \PrbCov{}{}{}{}{}{}_{#2;#3}
    }{}
    \ifthenelse{\equal{#1}{;_(;}}
    {
      \PrbCov{;_}{#2}{#3}{}{}{}\left(#4;#5\right)
    }{}
    \ifthenelse{\equal{#1}{;_(;=}}
    {
      \PrbCov{;_}{#2}{#3}{}{}{}\left(#4;#5\right)
    }{}
    \ifthenelse{\equal{#1}{;_(;;}}
    {
      \PrbCov{;_}{#2}{#3}{}{}{}\left(#4;#5;#6\right)
    }{}
    \ifthenelse{\equal{#1}{;_(;=;}}
    {
      \PrbCov{;_}{#2}{#3}{}{}{}\left(#4;#5;#6\right)
    }{}
    \ifthenelse{\equal{#1}{;(;}}
    {
      \PrbCov{}{}{}{}{}{}\left(#4;#5\right)
    }{}
    \ifthenelse{\equal{#1}{;(;;}}
    {
      \PrbCov{}{}{}{}{}{}\left(#4;#5;#6\right)
    }{}
  }
  \newcommand{\PrbCorr}[6]
  {
    \ifthenelse{\equal{#1}{}}
    {
      \ensuremath{\varrho}
    }{}
    \ifthenelse{\equal{#1}{(}}
    {
      \PrbCorr{}{}{}{}{}{}\left(#4\right)
    }{}
    \ifthenelse{\equal{#1}{(;}}
    {
      \PrbCorr{}{}{}{}{}{}\left(#4;#6\right)
    }{}
    \ifthenelse{\equal{#1}{_}}
    {
      \PrbCorr{}{}{}{}{}{}_{#2}
    }{}
    \ifthenelse{\equal{#1}{_(}}
    {
      \PrbCorr{}{}{}{}{}{}_{#2}\left(#4\right)
    }{}
    \ifthenelse{\equal{#1}{_(;}}
    {
      \PrbCorr{}{}{}{}{}{}_{#2}\left(#4;#6\right)
    }{}
    \ifthenelse{\equal{#1}{;_}}
    {
      \PrbCorr{}{}{}{}{}{}_{#2;#3}
    }{}
    \ifthenelse{\equal{#1}{;_(;}}
    {
      \PrbCorr{;_}{#2}{#3}{}{}{}\left(#4;#5\right)
    }{}
    \ifthenelse{\equal{#1}{;_(;=}}
    {
      \PrbCorr{;_}{#2}{#3}{}{}{}\left(#4;#5\right)
    }{}
    \ifthenelse{\equal{#1}{;_(;;}}
    {
      \PrbCorr{;_}{#2}{#3}{}{}{}\left(#4;#5;#6\right)
    }{}
    \ifthenelse{\equal{#1}{;_(;=;}}
    {
      \PrbCorr{;_}{#2}{#3}{}{}{}\left(#4;#5;#6\right)
    }{}
    \ifthenelse{\equal{#1}{;(;}}
    {
      \PrbCorr{}{}{}{}{}{}\left(#4;#5\right)
    }{}
    \ifthenelse{\equal{#1}{;(;;}}
    {
      \PrbCorr{}{}{}{}{}{}\left(#4;#5;#6\right)
    }{}
  }
  
  \newcommand*\prbS{\Omega}
  \newcommand*\prbs{\mathpzc{w}}
  \newcommand*\prbE{\mathscr{S}}
  
  \newcommand{\limit}[4]
  {
    \ifthenelse{\equal{#2}{>}}
    {
      \lim\limits_{#1 \searrow #3}\left( #4 \right)
    }
    {
      \ifthenelse{\equal{#2}{<}}
      {
        \lim\limits_{#1 \nearrow #3}\left( #4 \right)
      }
      {
        \lim\limits_{#1 \rightarrow #3}\left( #4 \right)
      }
    }
  }
  \newcommand*\dd{\mathop{}\!\mathrm{d}}
  
  \newcommand{\deriv}[3]
  {
    \ifthenelse{ \equal{#2}{}}
    {
      \ensuremath{\mathrm{D}_{#1}\left(#3\right)}
    }
    {
      \ensuremath{\mathrm{D}_{#1}^{#2}\left(#3\right)}
    }
  }
  \newcommand{\jacob}[2]
  {
    \frac{\partial(#1)}{\partial(#2)}
  }
  
  \newcommand{\varz}[3]
  {
    \ifthenelse{\equal{#1}{2}}
    {
      z_{#2},z_{#3}
    }{}
    \ifthenelse{\equal{#1}{Z2}}
    {
      Z_{#2},Z_{#3}
    }{}
    \ifthenelse{\equal{#1}{n}}
    {
      z_{#2},\dots,z_{#3}
    }{}
    \ifthenelse{\equal{#1}{Zn}}
    {
      Z_{#2},\dots,Z_{#3}
    }{}
    \ifthenelse{\equal{#1}{}}
    {
      z_{#2,#3}
    }{}
    \ifthenelse{\equal{#1}{Z}}
    {
      Z_{#2,#3}
    }{}
  }
  \newcommand{\vart}[3]
  {
    \ifthenelse{\equal{#1}{2}}
    {
      \theta_{#2},\theta_{#3}
    }{}
    \ifthenelse{\equal{#1}{T2}}
    {
      \vartheta_{#2},\vartheta_{#3}
    }{}
    \ifthenelse{\equal{#1}{n}}
    {
      \theta_{#2},\dots,\theta_{#3}
    }{}
    \ifthenelse{\equal{#1}{Tn}}
    {
      \vartheta_{#2},\dots,\vartheta_{#3}
    }{}
    \ifthenelse{\equal{#1}{}}
    {
      \theta_{#2,#3}
    }{}
    \ifthenelse{\equal{#1}{T}}
    {
      \vartheta_{#2,#3}
    }{}
  }
  \newcommand{\PrbPhysQ}[5] 
  {
    \ifthenelse{\equal{#1}{}}
    {
      p{\begin{xsmallmatrix}#4\\#2\end{xsmallmatrix}}
    }{}
    \ifthenelse{\equal{#1}{;}}
    {
      p{\begin{xsmallmatrix}#4 &;&#5\\#2 &;&#3\end{xsmallmatrix}}
    }{}
  }

  \newcommand{\braket}[3]
  {
    \ifthenelse{\equal{#2}{}}
    {
      \ensuremath{\left\langle#1 \vphantom{#3}\right| \left. #3 \vphantom{#1}\right\rangle}
    }{
      \ensuremath{\left\langle #1 \vphantom{#2#3} \right| #2 \left| #3 \vphantom{#1#2} \right\rangle}
    }
  } 

  \newcommand{\eqn}[3]    
  {
    \ifthenelse{\equal{#1}{}}
    {
      \ifthenelse{\equal{#2}{}}
      {
        \begin{eqnarray} #3 \end{eqnarray}
      }
      {
        \begingroup\def\arraystretch{#2} \begin{eqnarray} #3 \end{eqnarray}\endgroup
      }
    }
    {
      \ifthenelse{\equal{#2}{}}
      {
        \begingroup\arraycolsep=#1 \begin{eqnarray} #3 \end{eqnarray}\endgroup
      }
      {
        \begingroup\arraycolsep=#1 \def\arraystretch{#2} \begin{eqnarray} #3 \end{eqnarray}\endgroup
      }
    }
  }
  
  \newcommand{\eq}[2]   
  {
    \ifthenelse{ \equal{#1}{}}
    {
      {\begin{gather} #2 \end{gather}}
    }
    {
      {\begin{#1} #2 \end{#1}}
    }
  }
  
  \newcommand{\tab}[4]    
  {
    \ifthenelse{\equal{#1}{}}
    {
      \ifthenelse{\equal{#2}{}}
      {
        \begingroup\begin{array}{#3} #4 \end{array}\endgroup
      }
      {
        \begingroup\def\arraystretch{#2} \begin{array}{#3} #4 \end{array}\endgroup
      }
    }
    {
      \ifthenelse{\equal{#2}{}}
      {
        \begingroup\arraycolsep=#1 \begin{array}{#3} #4 \end{array}\endgroup
      }
      {
        \begingroup\arraycolsep=#1 \def\arraystretch{#2} \begin{array}{#3} #4 \end{array}\endgroup
      }
    }
  }
  \newcommand{\tabTexto}[4]    
  {
    \ifthenelse{\equal{#1}{}}
    {
      \ifthenelse{\equal{#2}{}}
      {
        \begin{tabular}{#3} #4 \end{tabular}
      }
      {
        \begingroup\def\arraystretch{#2} \begin{tabular}{#3} #4 \end{tabular}\endgroup
      }
    }
    {
      \ifthenelse{\equal{#2}{}}
      {
        \begingroup\arraycolsep=#1 \begin{tabular}{#3} #4 \end{tabular}\endgroup
      }
      {
        \begingroup\arraycolsep=#1 \def\arraystretch{#2} \begin{tabular}{#3} #4 \end{tabular}\endgroup
      }
    }
  }
  
\begin{document}

\preprint{APS/123-QED}

\title{Wigner and Bell Inequalities relationships and Kolmogorov's Axioms }

\author{Felipe  Andrade  Velozo}
\email{felipe.andrade.velozo@gmail.com}
\affiliation{Instituto  de  Ci\^encias  Sociais  Aplicadas  (ICSA).\\  Universidade  Federal  de  Alfenas  (UNIFAL),Campus  Avan\c{c}ado  de  Varginha-MG,  CEP  37048-395,  Brazil}

\author{Jos\'e  A.  C.  Nogales} 
\email{jnogales@dfi.ufla.br}
\affiliation{Departamento de F\'isica-DFI  e P\'os-Gradua\c{c}\~ao em Educa\c{c}\~ao Cient\'ifica e Ambiental Universidade Federal de Lavras (UFLA)  Caixa Postal 3037 - CEP 37200-900 - Lavras MG - Brasil}

\date{\today}

\begin{abstract}
In this work, we show that Bell's inequality violation of arise from the fact that the condition imposed upon the development of inequality is not respected when it is applied in the idealized experiment. Such a condition is that the quantities taken by probability must be non negative, and such a condition is represented by $|\Prb{(}{}{}{\prbs}{}{}|=\Prb{(}{}{}{\prbs}{}{}$. We will also show that, when trying to define the values of the joint probabilities of $ (Z_1, Z_2, Z_3) $, through the values obtained from the $ (Z_j, Z_k) $ pairs, we find that these values are negative, so not Kolmogorov's axiom is respected: $\Prb{(}{}{}{\prbs}{}{}\geq0$ in cases where Bell's inequality is violated, and we also show that only such violation is possible if Wigner's inequality, in a certain arrangement, is violated, and that both violations are related to the violation of one of Kolmogorov's axioms. At the end of the paper, we suggest a new interpretation of the probabilities involved, in order to avoid the situation of negative probabilities and the violation of Bell's inequality and, consequently, Wigner's inequality.

\end{abstract}

\maketitle

\section{Introduction}

In this work we explore the Bell like-inequalities over the strict probability theory of Kolmogorov. We start from a few straightforward hypotheses on probability as a physical property, and study some of their consequences. The main goal is to show that a coherent interpretation of probability offers an interesting focus  to understand  fundamental  problems of quantum mechanics. We show that the use of Kolmogorov's probability theory to describe results of quantum probability  experiments requires extreme care when different measurement outcomes are considered.
In spite of several claims that all the loopholes have been closed, according to recent experiments, we have no doubt that the violation of various Bell-type inequalities has been confirmed, and that in fact, perhaps no more Bell experiments are needed. In fact more recently a series of sophisticated Bell test experiments was realized  in 2015, the outcome of all experiments that violate a Bell inequality,  loophole-free Bell violation, was reported using entangled diamond spins over 1.3 km  \cite{Hensen2015} and corroborated by two experiments using entangled photon pairs.\cite{Giustina2015}, \cite{Christensen2013} However, we argued, that it is at list unattainable to perform a completely loophole-free Bell experiment. It was pointed out by several authors, that this results are no  complete, that all these inequalities are proven using  oversimplified probabilistic models which are inconsistent with the experimental protocols was used. A detailed discussion of the intimate relationship between experimental protocols and probabilistic models may be found anywhere.

The analysis of the probabilistic assumptions of Bell's arguments is extremely important for modern quantum physics and the consequences of the modern interpretation of the violation of Bell's inequality for the foundations of quantum mechanics are really relevant from a conceptual and practical reason. Hence, the conditions for deriving this inequality should be carefully checked. In previous work  we focus our considerations to strictly analyze the probabilistic conditions that have been assumed for the demonstration of CHSH inequality. Once the theoretical study of the basic assumptions for the CHSH inequality has been made, it is verified, through simulations, the manner in which the data of the samples should be used for such assumptions to be obeyed. In this way, both population and sample aspects shall be demonstrated. We conclude that Alain Aspect's experiment can be modeled by Classical Statistics in order to fully satisfy the CHSH inequality and that CHSH violation would only be possible if there is a violation of Kolmogorov's axioms\cite{Felipe2018}.

Now, in the communication we expand our proposal considering other inequalities like Bell, we will show that it is not necessary to consider any other hidden variable $\lambda$ again, besides the random variable $Z$, since it is irrelevant in the formalism, we are used and introduce the Wigner inequality an in the next we enlarge our analyses  once into a  conditional probabilities, in this point view we demonstrate that there are no violations of any inequalities therefore, there are no any conflict between Bell like inequalities and Kolmogorov theory. In very interesting case we observe that the solution of the system presents negative values for some probabilities negative probability values that have been widely employed as an indicator of the non-classically of quantum systems\cite{Ryu2019}. Since then, works that seek to establish a quantum probability or the use of other systems of probability axioms \cite{Khrennikov2008}, \cite{Pitowsky1989} can be found.

In the section \ref{sec:Explanation of the experiment}, we will start by presenting the probability function found in the Quantum Mechanics literature, associated with the experiment proposed in \cite {Bell1964} \cite{Wigner1970}.

In the section \ref{sec:Hidden variables} we will show that it is not necessary to consider any other variable $ \lambda $ besides the random variable $ Z $, making the analysis presented in \cite{Bell1964} unnecessary since it is irrelevant in the formalism they are used.

The inequality of Wigner \cite{Koc1993} \cite{Wigner1970} is presented in the section \ref{sec:Wigner inequality}, we merely emphasize the use of Komogorov's axioms in its development, because we will use this fact when associating Wigner's inequality Bell's inequality. We also made a little change, to use it when you don't have a triple of variables (since we defined the probability of pairs of random variables).

In the \ref{sec:Bell inequality} section, we also present Bell's inequality just to emphasize the importance of Komogorov's axioms in his statement. We developed and proposed another inequality, using the same logic of Bell's inequality, with only a small modification, where we consider any quantity $ Q $, which does not have the restrictions imposed by Kolmogorov's axioms, however we emphasize that it was necessary to use its absolute value $ | Q | $ in the statement. It is this quantity $ Q $ that is considered in the figure (\ref{fig:ExemploDesigBellFisQ}). 

In the section \ref{sec:Bell-Wigner and Kolmogorov} we show that when using Bell's inequality proposal, we obtain Wigner's inequality and, consequently, we relate the violation of Bell's inequality with the violation of Komogorov's axioms.

In the section \ref{sec:Bell in Quantum Mechanics} the use of the probabilities found in the Quantum Mechanics literature is shown, and we show in the figure (\ref{fig:DesigBellWignerFisQ}) the regions of violation of Bell's inequality compared to those of inequality of Wigner.
 
In the \ref{sec:Linear system} section we show that it is possible to determine probability values $ \Prb {_} {Z_1, Z_2, Z_3} {} {} {} {} $ following results from Quantum Mechanics, however
in cases where Bell's inequality is violated, as exemplified in the figure (\ref{fig:ExemploDesigBellFisQ}), we observe that the solution of the system presents negative values for some probabilities.

In the section \ref{Sec:Statistical estimators} we make, through the estimators  evident the need to reassess the use of the probabilities in Bell's inequality, once that it all came down to counting events. Inequality violations will only occur if $ n_ {j, k} <- n_ {j, k} $, but $n_{j,k}\in\set{N}{}{}$ .  We propose to consider the probability functions used as conditional probabilities.

In the section \ref{Loopholes} we show how to model the problem using conditional probabilities, showing how to calculate the expected values. In this modeling we expand the domain and the image of the random variables, in order to cover the possibilities in which there is no detection in any detector. In this modeling we demonstrate that there are no violations of inequality.

\begin{widetext}
\section{Explanation of the experiment}\label{sec:Explanation of the experiment}

In this section we set all the notation for the quantum inequalities. The experiment presented below explores the probabilities related to 2 particles, i.e.  a pair of photons, or a pair of electrons, and the interaction of each with the respective devices like  a polarizer, or a magnetic field gradient, respectively, each device is in the path of each particle, and after each device is a detector.

The possible results of the experiment are:

\begin{itemize}
  \item 
  in the case of the photon, or is absorbed (we will denote this result by $\prbs_{j,1}$) by the polarizer $j$ (placed in its path, and whose orientation can be rotated from an angle $\theta_j$ in the perpendicular plane your route) or cross it (we will denote it by $\prbs_{j,2}$). The physical amount (or property) of the photon that interacts with the polarizer, in this case, is its polarization (ie, the orientation of its electric field).
  
  \item
  in the case of an electron, or deviates its path in the opposite direction (denoted by $\prbs_{j,1}$) to the magnetic field of the $j$ apparatus (placed on its path, whose orientation can be rotated from an angle $\theta_j$ in the plane perpendicular to that of your route) or skew your trajectory towards the field (denoted by $\prbs_{j,2}$). The physical amount of the electron that interacts with the field in this case is its spin.
\end{itemize}

Thus, we have our sample space \cite{Doob1996}\cite{Kolmogorov1950} (set with all results) $\prbS_j=\{\prbs_{j,1},\prbs_{j,2}\}$ associated with the $j$ apparatus.

The random variable $Z$ associates with each result of $\prbS_j$ a real value, which in this case will be: $-1$ and $+1$. We will index the random variable to designate on which apparatus the event occurred, so $Z_j$ refers to the event on the apparatus $j$. We then have the following association: $Z_j(\prbs_{j,1})=-1$ and $Z_j(\prbs_{j,2})=+1$, so $\im{Z_j}=\{-1,+1\}$

The $\prbE_j$ event space will be the set of parts of $\prbS_j$, ie $\prbE_j=\Conj{P}{}{}(\prbS_j)=\{\emptyset,\{\prbs_1\},\{\prbs_2\},\prbS_j\}$. For the $Z_j$ random variable, we have the event space given by $\prbE_{Z_j}=\Conj{P}{}{}(\im{Z_j})=\{\emptyset,\{-1\},\{+1\},\{-1,+1\}\} $

For the pair of particles, we have the following event space: $\prbE_{j,k}=\Conj{P}{}{}(\prbS_j\times\prbS_k)$. For the random variables $(Z_j(w_j),Z_k(w_k))$ (where $w_j\in\{\prbs_{j,1},\prbs_{j, 2}\}$ and $ w_k\in\{\prbs_{k,1},\prbs_{k,2}\}$) we have the following sample space: $\prbE_{Z_j,Z_k}=\Conj{P}{}{}(\im{Z_j}\times\im{Z_k})$. There are two events of special interest in this space: the agreement, defined by $C:=\{(-1,-1), (+1,+1)\}$ and the disagreement, defined by $\bar C:=\{(-1,+1),(+1,-1)\}$ of the random variables $(Z_j(w_j),Z_k(w_k))$.

The joint probability\footnote{
  In the usual notation of statistics, in the index are random variables and in the argument of probability function are events and parameters, so 
  $\Prb{(}{}{}{\{\prbs\in\prbS;(Z(\prbs)\in C)\And(\theta=\theta_0)\}}{}{}
  \equiv\Prb{(;}{}{}{Z\in C}{}{\theta=\theta_0}
  \equiv\Prb{_(;}{Z}{}{C}{}{\theta_0}$. 
  Similarly, we have 
  $\Prb{(}{}{}{\{\prbs\in\prbS;(Z(\prbs)=z)\And(\theta=\theta_0)\}}{}{}
  \equiv\Prb{(;}{}{}{Z=z}{}{\theta=\theta_0}
  \equiv\Prb{_(;}{Z}{}{z}{}{\theta_0}$.
  The logical connective of disjunction (OR) is represented by  $ \Or $ symbol and the logical connective
  of conjunction (AND) is represented by $ \And $ symbol.
  For conditional probability we have 
  $\Prb{;_(;;}{W}{Z}{B}{C}{\theta_0}
  \equiv\frac{\Prb{(;}{}{}{W\in B}{}{\theta=\theta_0}}{\Prb{(;}{}{}{Z\in C}{}{\theta=\theta_0}}$ 
  and 
  $\Prb{;_(;}{W}{Z}{B}{C}{\theta_0}
  \equiv\frac{\Prb{(}{}{}{W\in B}{}{\theta=\theta_0}}{\Prb{(}{}{}{Z\in C}{}{\theta=\theta_0}}$. 
  Similarly, we have 
  $\Prb{;_(;;}{W}{Z}{w}{z}{\theta_0}
  \equiv\frac{\Prb{(;}{}{}{W=w}{}{\theta=\theta_0}}{\Prb{(;}{}{}{Z=z}{}{\theta=\theta_0}}$
  and 
  $\Prb{;_(;}{W}{Z}{w}{z}{\theta_0}
  \equiv\frac{\Prb{(}{}{}{W=w}{}{\theta=\theta_0}}{\Prb{(}{}{}{Z=z}{}{\theta=\theta_0}}$.
}
for 2 particles, whose properties are negatively correlated, is given by

\eq{}{\label{eq:BellProb2upla}
  \Prb{(;}{Z_j,Z_k}{}{Z_j=z_j,Z_k=z_k}{}{\vartheta_j,\vartheta_k}
  =
    \delim{\{}{}{
    \tab{0pt}{}{ll}{
      \tfrac{\sin^2(\vartheta_k-\vartheta_j)}{2}  &\fI(z_j,z_k)\in\{(-1,-1),(+1,+1)\}
      \\
      \tfrac{\cos^2(\vartheta_k-\vartheta_j)}{2}  &\fI(z_j,z_k)\in\{(-1,+1),(+1,-1)\}
    }
    }
  ,\  
  \delim{\{}{}{\tab{0pt}{}{l}{
  j\in\{1,2,3\}
  \\
  k\in\{1,2,3\}
  \\
  j\neq k
  }}
}

\noindent
therefore

\eqn{0pt}{}{\nonumber
  \Prb{(;}{}{}{Z_j=z,Z_k=z}{}{\vartheta_j,\vartheta_k}
  &=&
    \Prb{(;}{}{}{Z_j=-z,Z_k=-z}{}{\vartheta_j,\vartheta_k}
  \\\label{eq:BellProb2uplaRelacoes}
  &=&
    \tfrac{1}{2}-\Prb{(;}{}{}{Z_j=-z,Z_k=z}{}{\vartheta_j,\vartheta_k}
  \\\nonumber
  &=&
    \tfrac{1}{2}-\Prb{(;}{}{}{Z_j=z,Z_k=-z}{}{\vartheta_j,\vartheta_k}
  ,\quad  z\in\{-1,+1\}
}

\noindent
from which we find the marginal probability

\eq{}{\label{eq:BellProb1upla}
  \Prb{(}{Z_j}{}{Z_j=z_j}{}{\vartheta_j,\vartheta_k}
  =
    \sum_{z_k\in\{-1,+1\}}\delim{(}{)}{\Prb{(;}{Z_j,Z_k}{}{Z_j=z_j,Z_k=z_k}{}{\vartheta_j,\vartheta_k}}
  =
    \tfrac{1}{2},\quad z_j\in\{-1,+1\}
}

\noindent
and the conditional probability

\eqn{0pt}{}{\nonumber
  \Prb{;(;}{Z_j}{Z_k}{Z_j=z_j}{Z_k=z_k}{\vartheta_j,\vartheta_k}
  &=&
    \frac{
      \Prb{(;}{Z_j,Z_k}{}{Z_j=z_j,Z_k=z_k}{}{\vartheta_j,\vartheta_k}
    }{
      \Prb{(}{Z_k}{}{Z_k=z_k}{}{\vartheta_j,\vartheta_k}
    }
  \\\nonumber
  &=&
    \delim{\{}{}{
    \tab{0pt}{}{ll}{
      \frac{\sin^2(\vartheta_k-\vartheta_j)/2}{1/2}
      =\sin^2(\vartheta_k-\vartheta_j) &\fI(z_j,z_k)\in\{(-1,-1),(+1,+1)\}
      \\
      \frac{\cos^2(\vartheta_k-\vartheta_j)/2}{1/2}
      =\cos^2(\vartheta_k-\vartheta_j) &\fI(z_j,z_k)\in\{(-1,+1),(+1,-1)\}
    }}
}

We can see that Kolmogorov's axioms \cite{Kolmogorov1950} are satisfied for $\Prb{_}{Z_j,Z_k}{}{}{}{}$

\eqn{}{}{\nonumber
  \fa{j,k}{\fa{\vartheta_j,\vartheta_k}{\fa{z_j,z_k}{\Prb{(;}{Z_j,Z_k}{}{Z_j=z_j,Z_k=z_k}{}{\vartheta_j,\vartheta_k}\geq0}}}
  \\\nonumber
  \fa{j,k}{
    \fa{\vartheta_j,\vartheta_k}{
      \sum\limits_{z_j\in\{-1,+1\}}\delim{(}{)}{
        \sum\limits_{z_k\in\{-1,+1\}}\delim{}{}{
          \Prb{(;}{Z_j,Z_k}{}{Z_j=z_j,Z_k=z_k}{}{\vartheta_j,\vartheta_k}
        }
      }=1
    }
  }
}

Including from the last equation we get the following relation (where we adopt $C:=\{(-1,-1),(+1,+1)\}$, which is the agreement event and its complementary event $\bar C:=\{(-1,+1),(+1,-1)\}$)

\eq{}{\nonumber
  \Prb{_(}{Z_j,Z_k}{}{C}{}{}
  =
    1-\Prb{_(}{Z_j,Z_k}{}{\bar C}{}{}
  ,\ 
  C:=\{(-1,-1),(+1,+1)\},\ 
  \bar C:=\{(-1,+1),(+1,-1)\}
}

In case they are positively correlated particles, we have to $\vartheta_k-\vartheta_j$ must be replaced by $\tfrac{\pi}{2}-\delim{(}{)}{\vartheta_k-\vartheta_j}$. In the case of the photon experiment, we have to $\vartheta_j=\theta_j$, while in the case of electron experiments, we have to $\vartheta_j=\theta_j/2$.

From the probabilities we can calculate the expected value\footnote{
  In the statistical notation for expected values, we have that in the index is the random variable indicating which probability function is used, and in the argument, we have the function of the random variable, so we have $\PrbEsp{_(;}{Z}{}{f(Z)}{}{\theta_0}
\equiv\sum_{\prbs\in\prbS}\delim{(}{)}{f(Z(\prbs))\cdot\Prb{_(;}{Z}{}{Z(\prbs)}{}{\theta_0}}
\equiv\sum_{z\in\im {Z}}\delim{(}{)}{f(z)\cdot\Prb{_(;}{Z}{}{z}{}{\theta_0}}$
}

\eqn{}{}{\nonumber
  \PrbEsp{_(}{Z_j}{}{Z_j}{}{\vartheta_j,\vartheta_k}
  =
    \sum_{z\in\{-1,+1\}}\delim{(}{)}{z\cdot\Prb{(}{Z_j}{}{Z_j=z}{}{\vartheta_j,\vartheta_k}}
  =
    0
}

\noindent
and the expected value of the product $Z_j\cdot Z_k$

\eqn{0pt}{}{\nonumber
  \PrbEsp{_(}{Z_j,Z_k}{}{Z_j\cdot Z_k}{}{}
  &=&
    \sum_{\tab{0pt}{}{c}{z_j\in\{-1,+1\} \\ z_k\in\{-1,+1\}}}\delim{(}{)}{
      z_j\cdot z_k\cdot \Prb{(}{Z_j,Z_k}{}{Z_j=z_j,Z_k=z_k}{}{\vartheta_j,\vartheta_k}
    }
  \\\label{eq:BellEspProd}
  &=&
    \overunderbraces
    {
      &\br{3}{\scalebox{0.75}{$\Prb{_(}{Z_j,Z_k}{}{C}{}{}=$}}
      &
      &\br{3}{\scalebox{0.75}{$\Prb{_(}{Z_j,Z_k}{}{\bar C}{}{}=$}}}
    {
      &\Prb{_(}{Z_j,Z_k}{}{+1,+1}{}{\vartheta_j,\vartheta_k}
      &+
      &\Prb{_(}{Z_j,Z_k}{}{+1,+1}{}{\vartheta_j,\vartheta_k}
      &-
      &(\Prb{_(}{Z_j,Z_k}{}{+1,+1}{}{\vartheta_j,\vartheta_k}
      &+
      &\Prb{_(}{Z_j,Z_k}{}{+1,+1}{}{\vartheta_j,\vartheta_k})
      &
    }
    {}
  \\\nonumber
  &=&
    \delim{\{}{}{\tab{0pt}{}{l}{
      -\cos\delim{(}{)}{2\cdot(\vartheta_k-\vartheta_j)}
      \\
      2\cdot\Prb{_(}{Z_j,Z_k}{}{C}{}{}-1
      \\
      1-2\cdot\Prb{_(}{Z_j,Z_k}{}{\bar C}{}{}
    }}
}

\noindent
consequently, according to (\ref{eq:BellProb1upla}), we have the covariance\footnote{
  In the notation presented, we have 
  $\PrbCov{_(}{Z_j,Z_k}{}{f(Z_j,Z_k),g(Z_j,Z_k)}{}{}
  \equiv\PrbEsp{_(}{Z_j,Z_k}{}{f(Z_j,Z_k)\cdot g(Z_j, Z_k)}{}{}-\PrbEsp{_(}{Z_j,Z_k}{}{f(Z_j,Z_k)}{}{}\cdot\PrbEsp{_(}{Z_j,Z_k}{}{g(Z_j,Z_k)}{}{}$,
  so 
  $\PrbCov{_(}{Z_j,Z_k}{}{Z_j,Z_k}{}{}
  \equiv\PrbEsp{_(}{Z_j,Z_k}{}{Z_j\cdot Z_k}{}{}-\PrbEsp{_(}{Z_j,Z_k}{}{Z_j}{}{}\cdot\PrbEsp{_(}{Z_j,Z_k}{}{Z_k}{}{}$.
  In the notation present in quantum mechanics, we have 
  $\PrbEsp{_(}{Z_j,Z_k}{}{Z_j\cdot Z_k}{}{}
  \equiv\delim{\langle}{\rangle}{Z_j\cdot Z_k}$ 
  and 
  $\PrbEsp{_(}{Z_j,Z_k}{}{Z_j}{}{}
  \equiv\delim{\langle}{\rangle}{Z_j}$
}

\eqn{}{}{\nonumber
  \PrbCov{(}{Z_j,Z_k}{}{Z_j,Z_k}{}{\vartheta_j,\vartheta_k}
  \defeq
    \PrbEsp{(}{Z_j,Z_k}{}{Z_j\cdot Z_k}{}{}-\PrbEsp{(}{Z_j}{}{Z_j}{}{}\cdot\PrbEsp{(}{Z_k}{}{Z_k}{}{}
  =
    \PrbEsp{(}{Z_j,Z_k}{}{Z_j\cdot Z_k}{}{}
}

\noindent
in addition we have the conditional expected value

\eqn{0pt}{}{\nonumber
  \PrbEsp{;(;}{Z_j}{Z_k}{Z_j}{Z_k=z_k}{\vartheta_j,\vartheta_k}
  &=&
    \sum_{z_j\in\{-1,+1\}}\delim{(}{)}{
      z_j\cdot\Prb{;(;}{Z_j}{Z_k}{Z_j=z_j}{Z_k=z_k}{\vartheta_j,\vartheta_k}
    }
  \\\nonumber
  &=&
    \delim{\{}{}{
    \tab{0pt}{}{ll}{
      -\Prb{;(;}{Z_j}{Z_k}{Z_j=-1}{Z_k=-1}{\vartheta_j,\vartheta_k}+\Prb{;(;}{Z_j}{Z_k}{Z_j=+1}{Z_k=-1}{\vartheta_j,\vartheta_k}
      =-\cos(2\cdot(\vartheta_k-\vartheta_j))\fI z_k=-1
      \\
      -\Prb{;(;}{Z_j}{Z_k}{Z_j=-1}{Z_k=+1}{\vartheta_j,\vartheta_k}+\Prb{;(;}{Z_j}{Z_k}{Z_j=+1}{Z_k=+1}{\vartheta_j,\vartheta_k}
      =+\cos(2\cdot(\vartheta_k-\vartheta_j))\fI z_k=+1
    }}
}

It is important to note that neither time nor position, in which the experiments take place, was taken into account at any time (this is easily observed by finding that the probabilities do not depend on time or space). Therefore, we are paying attention only to the statistical question, not considering the relativistic questions of the experiment (regarding signal transmission). Thus, it is not considered the moment in which the particles correlate with each other, neither the moment nor the place where the measurement occurs, is only considered the result of the experiment. For such considerations to exist, there should be in the expression that gives us the probability, the variables (parameters of the probability function) related to the moments at which events occur (the correlation related and the measurement related) and some distance related variable. between the particles (or the path taken to separate them) at the time of measurement.

\section{Hidden variables}\label{sec:Hidden variables}

In probability theory, what we need is to define the event space, to define from it the probability function. Random variables are just functions that assign certain values to the possible results $\prbs$, and are useful for establishing a relationship between Borel's $\sigma$-algebra and the $\sigma$-algebra generated by the random variable. Parameters are part of the definition of the probability function, so with each value set for the parameters we have a new probability function.

Thus, the random variable only links a result to a number (if it is multidimensional, it links to a sequence of numbers), so defining it by no means implies that such a result will occur. Therefore, the random variable has to be well defined since it concerns the results (which must be recognizable).

If we want some event to be certain to occur somehow, we have to have the probability function associating that event with the value 1 (ie 100\%). One way to do this can be through a certain value, or a sequence of values, assigned to the parameters. An example is the random walk (Wiener process), which for $t=0$ we have that the particle position is certainly equal to 0.

Let's begin by defining the random variable as it is defined in the articles on hidden variables, where we have

\eqn{}{}{\nonumber
  \tab{0pt}{}{rcl}{
    Z_j:\prbS\times\Theta&\rightarrow&\{-1,+1\}
    \\
    (\lambda,\vartheta)&\mapsto &Z_{j}(\lambda,\vartheta)
  }
}

\noindent
therefore, $\lambda$ plays the same role as $\prbs$ ($Z(\lambda,\vartheta)\equiv Z(\prbs,\vartheta)$), where each possible result obtained from the experiment associates a real number. We will now see that the consideration made in \cite{Bell1964} about the hidden variable $\lambda$ is unnecessary.

Let's see how the consideration about $\lambda$ being a continuous variable does not contribute anything. Let's start by calculating the expected value, as proposed in \cite{Bell1964} we have

\eqn{}{}{\label{eq:EspProdFisQVarOculta}
  \PrbEsp{_(}{\lambda}{}{Z_{j}(\lambda,\vartheta_j)\cdot Z_{k}(\lambda,\vartheta_k)}{}{}
  =
    \int_{\lambda\in\prbS_{j,k}}
    \delim{(}{)}{Z_j(\lambda,\vartheta_j)\cdot Z_k(\lambda,\vartheta_k)\cdot\Prb{d(}{}{}{\lambda}{}{}}\dd\lambda
    ,\quad j\neq k
}

\noindent
We can partition $\prbS_{j,k}$ as follows

\eqn{}{}{\nonumber
  \Lambda_u^v:=\{\lambda\in\prbS_{j,k};(Z_j(\lambda,\vartheta)=u)\And(Z_k(\lambda,\vartheta)=v)\}
}

\noindent
so if $\lambda\in\Lambda_{-1}^{+1}$ then $\prbS_{j,k}$, where $Z_j(\lambda,\vartheta)=-1$ and $Z_k(\lambda,\vartheta)=+1$. Obviously, $\Lambda_u^v$ are disjoint, so our expected value can be rewritten like this.

\eqn{0pt}{}{\nonumber
  \PrbEsp{_(}{\lambda}{}{Z_{j}(\lambda,\vartheta_j)\cdot Z_{k}(\lambda,\vartheta_k)}{}{}
  &=&
    \sum_{u\in\{-1,+1\}}\delim{(}{)}{
      \sum_{v\in\{-1,+1\}}\delim{(}{)}{
        \int_{\lambda\in\Lambda_{u}^{v}}
        \delim{(}{)}{Z_j(\lambda,\vartheta_j)\cdot Z_k(\lambda,\vartheta_k)\cdot\Prb{d(}{}{}{\lambda}{}{}}\dd\lambda
      }
    }
  \\\nonumber
  &=&
    \sum_{u\in\{-1,+1\}}\delim{(}{)}{
      \sum_{v\in\{-1,+1\}}\delim{(}{)}{
        \int_{\lambda\in\Lambda_{u}^{v}}
        \delim{(}{)}{u\cdot v\cdot\Prb{d(}{}{}{\lambda}{}{}}\dd\lambda
      }
    }
  \\\nonumber
  &=&
    \sum_{u\in\{-1,+1\}}
    \delim{(}{}{\vphantom{\sum_{v\in\{-1,+1\}}}}
      \sum_{v\in\{-1,+1\}}
      \delim{(}{}{\vphantom{\int_{\lambda\in\Lambda_{u}^{v}}}}
        u\cdot v\cdot
        \underbrace{
          \int_{\lambda\in\Lambda_{u}^{v}}
          \delim{(}{)}{\Prb{d(}{}{}{\lambda}{}{}}\dd\lambda
        }_{
          =\Prb{(;}{Z_j,Z_k}{}{Z_j=u,Z_k=v}{}{\vartheta_j,\vartheta_k}
        }
      \delim{}{)}{\vphantom{\int_{\lambda\in\Lambda_{u}^{v}}}}
    \delim{}{)}{\vphantom{\sum_{v\in\{-1,+1\}}}}
  \\\nonumber
  &=&
    \sum_{u\in\{-1,+1\}}\delim{(}{)}{
      \sum_{v\in\{-1,+1\}}\delim{(}{)}{
        u\cdot v\cdot\Prb{(;}{Z_j,Z_k}{}{Z_j=u,Z_k=v}{}{\vartheta_j,\vartheta_k}
      }
    }
  \\\nonumber
  &=&
    \PrbEsp{_(}{Z_j,Z_k}{}{Z_j(\prbs)\cdot Z_k(\prbs)}{}{}
}

\noindent
Therefore, we can conclude that the consideration that $\lambda$ is continuous (\ref{eq:EspProdFisQVarOculta}) is unnecessary, because in the end what matters is the values of the discrete variables $Z_j$ and $Z_k$.

Also, we have that any dependency on the parameters ($\vartheta_j$ and $\vartheta_k$) that could have the random variables $Z_j$ and $Z_j$ become irrelevant since only the values of the random variables is that appear, all reliance on parameters being restricted to the probability function $\Prb{_}{Z_j,Z_k}{}{}{}{}$. So even if we considered a $Z_j$ dependency on $\vartheta_k$, it would all boil down to $Z_j\in\{-1,+1\}$ (ie we only need the $Z_j$ image) . For this reason, in probability theory, we have to consider as a triple probability space $(\prbS,\prbE,\Prb{}{}{}{}{}{})$

\section{Wigner inequality}\label{sec:Wigner inequality}

The Wigner's inequality \cite{Wigner1970}, as we will see below, assumes that there is a probability function 
$\Prb{(}{Z_1,Z_2,Z_3}{}{Z_1,Z_2,Z_3}{}{}$ defined for the triple of random variables $(Z_1,Z_2,Z_3)$. Each of the variables has two values in its image, that is, $Z_j(\prbs)\in\{z_j,\bar z_j\}$ to $j\in\{1,2,3\}$. Thus, using the marginal probability property $\Prb{(}{j,k}{}{Z_j=z_j,Z_k=z_k}{}{}=\Prb{(}{j,k,l}{}{Z_j=z_j,Z_k=z_k,Z_l=z_l}{}{}+\Prb{(}{j,k,l}{}{Z_j=z_j,Z_k=z_k,Z_l=\bar z_l}{}{}$ we have

\eqn{0pt}{}{\nonumber
  \Prb{(}{Z_j,Z_k}{}{z_j,z_k}{}{}+\Prb{(}{Z_k,Z_l}{}{\bar z_k,z_l}{}{}
  &=&
    \Prb{(}{Z_j,Z_k,Z_l}{}{z_j,z_k,\bar z_l}{}{}
    +
    \underbrace{
      \Prb{(}{Z_j,Z_k,Z_l}{}{z_j,z_k,z_l}{}{}+\Prb{(}{Z_j,Z_k,Z_l}{}{z_j,\bar z_k,z_l}{}{}
    }_{=\Prb{(}{Z_j,Z_l}{}{z_j,z_l}{}{}}
    +
    \Prb{(}{Z_j,Z_k,Z_l}{}{\bar z_j,\bar z_k,z_l}{}{}
  \\\nonumber
  &=&
    \underbrace{\Prb{(}{Z_j,Z_k,Z_l}{}{z_j,z_k,\bar z_l}{}{}}_{\geq0}
    +\Prb{(}{Z_j,Z_l}{}{z_j,z_l}{}{}
    +\underbrace{\Prb{(}{Z_j,Z_k,Z_l}{}{\bar z_j,\bar z_k,z_l}{}{}}_{\geq0}
}

\noindent
and from Kolmogorov's axioms ($\fa{A\in\prbE}{\Prb{(}{}{}{A}{}{}\geq0}$) we find Wigner's inequality

\eq{}{\label{eq:DesigWigner}
  \underbrace{
    \Prb{(}{j,k}{}{Z_j=z_j,Z_k=z_k}{}{}+\Prb{(}{k,l}{}{Z_k=\bar z_k,Z_l=z_l}{}{}-\Prb{(}{j,l}{}{Z_j=z_j,Z_l=z_l}{}{}
  }_{=\Prb{(}{}{}{Z_j=z_j,Z_k=z_k,Z_l=\bar z_l}{}{}+\Prb{(}{}{}{Z_j=\bar z_j,Z_k=\bar z_k,Z_l=z_l}{}{}}
  \geq0
}

For such inequality, we only assume that there is $\Prb{(}{Z_1,Z_2,Z_3}{}{Z_1,Z_2,Z_3}{}{}$ and that images of random variables are given by a set of two values $\{z_j,\bar z_j\}$, however, we can extend these assumptions by simply requiring $z_j$ to be a possible result set (a sample space event) and $\bar z_j$ to be its respective complementary set $\complement_{\prbS}(z_j)$, resulting in

\eqn{0pt}{}{\nonumber
  &&\Prb{(}{\mathbf{Z}}{}{\mathbf{Z}\in z_1\cap z_2}{}{}
  +\Prb{(}{\mathbf{Z}}{}{\mathbf{Z}\in\bar z_2\cap z_3}{}{}
  =
  \\\nonumber
  &&=
    \Prb{(}{\mathbf{Z}}{}{\mathbf{Z}\in z_1\cap z_2\cap \bar z_3}{}{}
    +
    \underbrace{
      \Prb{(}{\mathbf{Z}}{}{\mathbf{Z}\in z_1\cap z_2\cap z_3}{}{}
      +\Prb{(}{\mathbf{Z}}{}{\mathbf{Z}\in z_1\cap\bar z_2\cap z_3}{}{}
    }_{=\Prb{(}{\mathbf{Z}}{}{\mathbf{Z}\in z_1\cap z_3}{}{}}
    +\Prb{(}{\mathbf{Z}}{}{\mathbf{Z}\in \bar z_1\cap\bar z_2\cap z_3}{}{}
  =\\\nonumber
  &&=
    \underbrace{\Prb{(}{\mathbf{Z}}{}{\mathbf{Z}\in z_1\cap z_2\cap\bar z_3}{}{}}_{\geq0}
    +\Prb{(}{\mathbf{Z}}{}{\mathbf{Z}\in z_1\cap z_3}{}{}
    +\underbrace{\Prb{(}{\mathbf{Z}}{}{\mathbf{Z}\in\bar z_1\cap\bar z_2\cap z_3}{}{}}_{\geq0}
  \geq\\\nonumber
  &&\geq
    \Prb{(}{\mathbf{Z}}{}{\mathbf{Z}\in z_1\cap z_3}{}{}
}

\noindent
where we use the fact that $\fa{j}{z_j\subseteq\prbS}$, $z_j\cup\bar z_j=\prbS$ and that $z_j\cap\bar z_j=\emptyset$, and one of Kolmogorov's axioms (which establishes that the probability of joining disjoint events is equal to the sum of the probabilities)

\eqn{0pt}{}{\nonumber
  \Prb{(}{\mathbf{Z}}{}{\mathbf{Z}\in z_j\cap z_k\cap z_l}{}{}
  +\Prb{(}{\mathbf{Z}}{}{\mathbf{Z}\in z_j\cap z_k\cap\bar z_l}{}{}
  &=&
    \Prb{(}{\mathbf{Z}}{}{\mathbf{Z}\in\delim{(}{)}{z_j\cap z_k\cap z_l}\cup\delim{(}{)}{z_j\cap z_k\cap\bar z_l}}{}{}
  =\\\nonumber
  &=&
    \Prb{(}{\mathbf{Z}}{}{\mathbf{Z}\in
      \overbrace{
        z_j\cap z_k\cap\underbrace{(z_l\cup\bar z_l)}_{=\prbS}
      }^{z_j\cap z_k=}
    }{}{}
  =
    \Prb{(}{\mathbf{Z}}{}{\mathbf{Z}\in z_j\cap z_k}{}{}
}

This inequality serves not only for $\mathbf{Z}=(Z_1,Z_2,Z_3)$ but also when $\mathbf{Z}$ is a univariate, or multivariate random variable with a dimension other than 3.

We note, therefore, that Wigner's inequality is only violated if Kolmogorov's axiom is violated, that is, if $\ex{A}{\Prb{(}{Z}{}{\mathbf{Z}\in A}{}{}<0}$.

\section{Bell inequality}\label{sec:Bell inequality}

The Bell's inequality \cite{Bell1964} assumes that there is a joint probability function $\Prb{(}{Z_j,Z_k,Z_l}{}{Z_j,Z_k,Z_l}{}{}$, defined such way that can be obtained the marginal probabilities $\Prb{(}{Z_j,Z_k}{}{Z_j,Z_k}{}{}$ related to the probabilities of two correlated particles (given above). We will see that it will not always be possible to find such a probability function.

We will begin by demonstrating a basic inequality, from which we will obtain Bell's inequality. Let the variables be random

\eqn{0pt}{}{\nonumber
  Z_j:\prbS_j&\rightarrow&\{-1,+1\}
  \\\nonumber
  Z_j:\prbs&\mapsto&Z_j(\prbs)
}

\noindent
so we have to $\delim{|}{|}{Z_j}\equiv 1\equiv\delim{(}{)}{Z_j}^2$ and $1\pm Z_j\in\{0,+2\}$ (we are omitting the $\prbs$ arguments to simplify writing). Using these results, we have

\eq{}{\nonumber
  Z_j\pm Z_k
  \equiv 
    Z_j\pm\underbrace{(Z_j)^2}_{=1}\cdot Z_k
  \equiv
    Z_j\cdot(1\pm Z_j\cdot Z_k)
}

\noindent
calculating the absolute value we have

\eq{}{\label{eq:DesigBellVariaveis}
  |Z_j\pm Z_k|
  \equiv
    \underbrace{|Z_j|}_{=1}\cdot|\underbrace{1\pm Z_j\cdot Z_k}_{\in\{0,+2\}}|
  \equiv
    1\pm Z_j\cdot Z_k
}

\noindent
multiplying (\ref{eq:DesigBellVariaveis}) by 
$\Prb{(}{}{}{\prbs}{}{}:=\Prb{(}{}{}{Z_j(\prbs),Z_k(\prbs),Z_l(\prbs)}{}{}$, we have

\eq{}{\nonumber
  \delim{|}{|}{Z_j\pm Z_k}\cdot\Prb{(}{}{}{\prbs}{}{}
  =
    \delim{(}{)}{1\pm Z_j\cdot Z_k}\cdot\Prb{(}{}{}{\prbs}{}{}
}

\noindent
and calculating the expected value $\PrbEsp{}{Z_j,Z_k}{}{}{}{}$, we have

\eqn{0pt}{}{\nonumber
  \underbrace{
    \sum\limits_{\prbs\in\prbS}\delim{(}{)}{\Prb{(}{}{}{\prbs}{}{}}
  }_{=1}
  \pm\underbrace{
    \sum\limits_{\prbs\in\prbS}\delim{(}{)}{Z_j\cdot Z_k\cdot\Prb{(}{}{}{\prbs}{}{}}
  }_{\PrbEsp{(}{}{}{Z_j\cdot Z_k}{}{}}
  &=&
    \overbrace{
      \sum\limits_{\prbs\in\prbS}\delim{(}{)}{\delim{|}{|}{Z_j\pm Z_k}\cdot\Prb{(}{}{}{\prbs}{}{}}
      =
        \sum\limits_{\prbs\in\prbS}\delim{(}{)}{\delim{|}{|}{Z_j\pm Z_k}\cdot\delim{|}{|}{\Prb{(}{}{}{\prbs}{}{}}}
    }^{\fa{\prbs\in\prbS}{\Prb{(}{}{}{\prbs}{}{}\geq0}\If}
  =\\\label{eq:DesigBellDemo}
  &=&
    \sum\limits_{\prbs\in\prbS}\delim{|}{|}{\delim{(}{)}{Z_j\pm Z_k}\cdot\Prb{(}{}{}{\prbs}{}{}}
  \geq\\\nonumber
  &\geq&
    \Bigg|
    \underbrace{
      \sum\limits_{\prbs\in\prbS}\delim{(}{)}{Z_j\cdot\Prb{(}{}{}{\prbs}{}{}}
    }_{=\PrbEsp{(}{}{}{Z_j}{}{}}
    \pm
    \underbrace{
      \sum\limits_{\prbs\in\prbS}\delim{(}{)}{Z_k\cdot\Prb{(}{}{}{\prbs}{}{}}
    }_{=\PrbEsp{(}{}{}{Z_k}{}{}}
    \Bigg|
}

\eq{}{\nonumber
  \PrbEsp{(}{Z_j,Z_k}{}{1\pm Z_j\cdot Z_k}{}{}
  \geq
    \delim{|}{|}{\PrbEsp{(}{Z_j,Z_k}{}{Z_j}{}{}\pm\PrbEsp{(}{Z_j,Z_k}{}{Z_k}{}{}}
}

When making the following substitutions $Z_j:=Z_1\cdot Z_2$ and $Z_k:=Z_2\cdot Z_3$, where $Z_j\cdot Z_k\equiv Z_1\cdot(Z_2)^2\cdot Z_3\equiv Z_1\cdot Z_3$, we find Bell's inequality
 
\eq{}{\label{eq:DesigBell-}
  1-\PrbEsp{(}{Z_1,Z_2,Z_3}{}{Z_1\cdot Z_3}{}{}
  \geq
    \delim{|}{|}{\PrbEsp{(}{Z_1,Z_2,Z_3}{}{Z_1\cdot Z_2}{}{}-\PrbEsp{(}{Z_1,Z_2,Z_3}{}{Z_2\cdot Z_3}{}{}}
}

\noindent
analogously we also get

\eq{}{\nonumber
  1+\PrbEsp{(}{Z_1,Z_2,Z_3}{}{Z_1\cdot Z_3}{}{}
  \geq
    \delim{|}{|}{\PrbEsp{(}{Z_1,Z_2,Z_3}{}{Z_1\cdot Z_2}{}{}+\PrbEsp{(}{Z_1,Z_2,Z_3}{}{Z_2\cdot Z_3}{}{}}
}

\noindent
where $\PrbEsp{(}{}{}{Z_j\cdot Z_k}{}{}\equiv\PrbEsp{_(}{Z_1,Z_2,Z_2}{}{Z_j\cdot Z_k}{}{}\equiv
\sum_{(z_1,z_2,z_3)\in\{-1,+1\}^3}\delim{(}{)}{z_j\cdot z_k\cdot\Prb{(}{}{}{Z_1=z_1,Z_2=z_2,Z_3=z_3}{}{}}$
and $(j,k)\in\{(1,2),(1,3),(2,3)\}$.

It is important to note that in the demonstration we made use of Kolmogorov's axioms ($\fa{\prbs\in\prbS}{\Prb{(}{}{}{\prbs}{}{}\geq0}$ and $\sum_{\prbs\in\prbS}(\Prb{(}{}{}{\prbs}{}{})=1$) to get to Bell's inequalities.

We can change the statement, by exchanging $\Prb{(}{}{}{\prbs}{}{}$ for another quantity $Q(\prbs)$ (dependent or not of $\prbs$) in (\ref{eq:DesigBellVariaveis}), which can even assume negative values or greater than 1. Thus we have

\eq{}{\nonumber
  \delim{|}{|}{Z_j\pm Z_k}\cdot|Q(\prbs)|
  =
    \delim{(}{)}{1\pm Z_j\cdot Z_k}\cdot|Q(\prbs)|
}

\eqn{0pt}{}{\nonumber
  \sum\limits_{\prbs\in\prbS}\delim{(}{)}{\delim{(}{)}{1\pm Z_j\cdot Z_k}\cdot|Q(\prbs)|}
  &=&
    \sum\limits_{\prbs\in\prbS}\delim{(}{)}{|Q(\prbs)|}
    \pm\sum\limits_{\prbs\in\prbS}\delim{(}{)}{Z_j\cdot Z_k\cdot|Q(\prbs)|}
  =\\\nonumber
  &=&
    \sum\limits_{\prbs\in\prbS}\delim{(}{)}{\delim{|}{|}{Z_j\pm Z_k}\cdot\delim{|}{|}{Q(\prbs)}}
  =
    \sum\limits_{\prbs\in\prbS}\delim{|}{|}{\delim{(}{)}{Z_j\pm Z_k}\cdot Q(\prbs)}
  \geq\\\nonumber
  &\geq&
    \bigg|
      \sum\limits_{\prbs\in\prbS}\delim{(}{)}{Z_j\cdot Q(\prbs)}
      \pm\sum\limits_{\prbs\in\prbS}\delim{(}{)}{Z_k\cdot Q(\prbs)}
    \bigg|
}

\noindent
so we have the following inequality for any quantity $Q$

\eqn{0pt}{}{\label{eq:DesigBellGeral}
  \sum\limits_{\prbs\in\prbS}\delim{(}{)}{|Q(\prbs)|}
  \pm\sum\limits_{\prbs\in\prbS}\delim{(}{)}{Z_j\cdot Z_k\cdot|Q(\prbs)|}
  &\geq&
    \bigg|
      \sum\limits_{\prbs\in\prbS}\delim{(}{)}{Z_j\cdot Q(\prbs)}
      \pm\sum\limits_{\prbs\in\prbS}\delim{(}{)}{Z_k\cdot Q(\prbs)}
    \bigg|
}

such inequality will be useful later on, when we meet the "probabilities" of $(Z_1,Z_2,Z_3)$.

\subsection{Relationship of Bell's inequality used in the quantum mechanics with the Kolmogorov
probability axioms}\label{sec:Bell-Wigner and Kolmogorov}

In the quantum mechanics literature, we find that

\eq{}{\nonumber
  \PrbEsp{_(}{Z_j,Z_k}{}{Z_j\cdot Z_k}{}{}
  =-\cos(2\cdot(\vartheta_k-\vartheta_j))
  =2\cdot\Prb{(}{j,k}{}{(Z_j,Z_k)\in C}{}{}-1
}

\noindent
and based on that, even if we don't have $\Prb{_}{Z_1,Z_2,Z_3}{}{}{}{}$, we find the following assignment

\eq{}{\nonumber
  \PrbEsp{_(}{Z_1,Z_2,Z_3}{}{Z_j\cdot Z_k}{}{}
  \equiv
    \PrbEsp{_(}{Z_j,Z_k}{}{Z_j\cdot Z_k}{}{}
}

In probability theory, since before arriving at such a formula, it is firstly assumed that the joint probability function $\Prb{_}{Z_1,Z_2,Z_3}{}{}{}{}$ (properly determining the sample space and the event space). Such care ensures the validity of the formula (as they will conform to Kolmogorov's axioms). We will see later how this simple assignment will lead to noncompliance with one of Kolmmogorov's axioms.

Performing the substitution (\ref{eq:BellEspProd}) on Bell's inequality (\ref{eq:DesigBell-}), we have

\eq{}{\label{eq:RelacaoBellWigner}
  \overunderbraces
  {
    &\br{3}{\scalebox{0.75}{$\Prb{_(}{Z_2,Z_3}{}{C}{}{}
    -\Prb{_(}{Z_1,Z_2}{}{C}{}{}+1-\Prb{_(}{Z_1,Z_3}{}{C}{}{}\geq0$}}
  }
  {
    &1-\delim{(}{)}{2\cdot\Prb{_(}{Z_1,Z_3}{}{C}{}{}-1}
    &\geq
    &
      (2\cdot\Prb{_(}{Z_1,Z_2}{}{C}{}{}-1)-(2\cdot\Prb{_(}{Z_2,Z_3}{}{C}{}{}-1)
    &\geq
    &
      -(1-\delim{(}{)}{2\cdot\Prb{_(}{Z_1,Z_3}{}{C}{}{}-1})&
  }
  {
    &
    &
    &\br{3}{\scalebox{0.75}{$\Prb{_(}{Z_1,Z_2}{}{C}{}{}
    -\Prb{_(}{Z_2,Z_3}{}{C}{}{}+1-\Prb{_(}{Z_1,Z_3}{}{C}{}{}{}\geq0$}}
  }
}

\noindent
resulting in two other inequalities, both identical to Wigner's inequality (\ref{eq:DesigWigner})

\eqn{0pt}{}{\nonumber
  &&\overunderbraces{
    &\br{2}{\scalebox{0.75}{$\Prb{_(}{Z_1,Z_2}{}{\bar C}{}{}=$}}
  }{
    \Prb{_(}{Z_2,Z_3}{}{C}{}{}
    &-\Prb{_(}{Z_1,Z_2}{}{C}{}{}+
    &1
    &-\Prb{_(}{Z_1,Z_3}{}{C}{}{}
    &\geq0
  }{
    &
    &\br{2}{\scalebox{0.75}{$=\Prb{_(}{Z_1,Z_3}{}{\bar C}{}{}$}}
  }
  \If\\\nonumber
  &&\If
  \delim{\{}{}{\tab{0pt}{}{l}{
    \overbrace{
      \Prb{_(}{Z_2,Z_3}{}{+,+}{}{}+\Prb{_(}{Z_1,Z_2}{}{+,-}{}{}-\Prb{_(}{Z_1,Z_3}{}{+,+}{}{}
    }^{\Prb{_(}{Z_1,Z_2,Z_3}{}{+,-,-}{}{}+\Prb{_(}{Z_1,Z_2,Z_3}{}{-,+,+}{}{}=}
    +\overbrace{
      \Prb{_(}{Z_2,Z_3}{}{-,-}{}{}+\Prb{_(}{Z_1,Z_2}{}{-,+}{}{}-\Prb{_(}{Z_1,Z_3}{}{-,-}{}{}
    }^{\Prb{_(}{Z_1,Z_2,Z_3}{}{+,-,-}{}{}+\Prb{_(}{Z_1,Z_2,Z_3}{}{-,+,+}{}{}=}
    \geq0
    \\
    \underbrace{
      \Prb{_(}{Z_2,Z_3}{}{+,+}{}{}-\Prb{_(}{Z_1,Z_2}{}{+,+}{}{}+\Prb{_(}{Z_1,Z_3}{}{+,-}{}{}
    }_{=\Prb{_(}{Z_1,Z_2,Z_3}{}{+,-,-}{}{}+\Prb{_(}{Z_1,Z_2,Z_3}{}{-,+,+}{}{}}
    +\underbrace{
      \Prb{_(}{Z_2,Z_3}{}{-,-}{}{}-\Prb{_(}{Z_1,Z_2}{}{-,-}{}{}+\Prb{_(}{Z_1,Z_3}{}{-,+}{}{}
    }_{=\Prb{_(}{Z_1,Z_2,Z_3}{}{+,-,-}{}{}+\Prb{_(}{Z_1,Z_2,Z_3}{}{-,+,+}{}{}}
    \geq0
  }}
}

\noindent
and

\eqn{0pt}{}{\nonumber
  &&\overunderbraces{
    &\br{2}{\scalebox{0.75}{$\Prb{_(}{Z_2,Z_3}{}{\bar C}{}{}=$}}
  }{
    \Prb{_(}{Z_1,Z_2}{}{C}{}{}&-\Prb{_(}{Z_2,Z_3}{}{C}{}{}+&1&-\Prb{_(}{Z_1,Z_3}{}{C}{}{}{}&\geq0
  }{
    &&\br{2}{\scalebox{0.75}{$=\Prb{_(}{Z_1,Z_3}{}{\bar C}{}{}$}}
  }
  \If\\\nonumber
  &&\If
  \delim{\{}{}{\tab{0pt}{}{l}{
    \overbrace{
      \Prb{_(}{Z_1,Z_2}{}{+,+}{}{}+\Prb{_(}{Z_2,Z_3}{}{-,+}{}{}
      -\Prb{_(}{Z_1,Z_3}{}{+,+}{}{}
    }^{
      \Prb{_(}{Z_1,Z_2,Z_3}{}{+,+,-}{}{}+\Prb{_(}{Z_1,Z_2,Z_3}{}{-,-,+}{}{}=
    }
    +\overbrace{
      \Prb{_(}{Z_1,Z_2}{}{-,-}{}{}+\Prb{_(}{Z_2,Z_3}{}{+,-}{}{}
      -\Prb{_(}{Z_1,Z_3}{}{-,-}{}{}
    }^{
      \Prb{_(}{Z_1,Z_2,Z_3}{}{+,+,-}{}{}+\Prb{_(}{Z_1,Z_2,Z_3}{}{-,-,+}{}{}=
    }
    \geq0
    \\
    \underbrace{
      \Prb{_(}{Z_1,Z_2}{}{+,+}{}{}-\Prb{_(}{Z_2,Z_3}{}{+,+}{}{}+\Prb{_(}{Z_1,Z_3}{}{-,+}{}{}
    }_{
      =\Prb{_(}{Z_1,Z_2,Z_3}{}{+,+,-}{}{}+\Prb{_(}{Z_1,Z_2,Z_3}{}{-,-,+}{}{}
    }
    +\underbrace{
      \Prb{_(}{Z_1,Z_2}{}{-,-}{}{}-\Prb{_(}{Z_2,Z_3}{}{-,-}{}{}+\Prb{_(}{Z_1,Z_3}{}{+,-}{}{}
    }_{
      =\Prb{_(}{Z_1,Z_2,Z_3}{}{+,+,-}{}{}+\Prb{_(}{Z_1,Z_2,Z_3}{}{-,-,+}{}{}
    }
    \geq0
  }}
}

Since we have linked Bell's inequality with Wigner's inequality (when $(Z_1,Z_2,Z_3)=(+1,-1,+1)$ is replaced with Wigner inequality (\ref{eq:DesigWigner}), we get the previous result), too we have a relationship with Kolmogorov's axiom. We then conclude that Bell's inequality is not satisfied if the Kolmogorov's axiom ($\fa{A}{\Prb{(}{}{}{A}{}{}\geq0}$) is not satisfied.

\subsection{Use of Bell's inequality in the literature of quantum mechanics}\label{sec:Bell in Quantum Mechanics}

Now we will proceed by performing the replacement found in the quantum mechanics literature 
(\ref{eq:BellProb2upla})(\ref{eq:BellProb2uplaRelacoes}), where

\eq{}{\nonumber
  \Prb{_(}{Z_j,Z_k}{}{+,+}{}{}
  =
    \tfrac{1}{2}\cdot\sin\delim{(}{)}{\theta_k-\theta_j}
  =
    \tfrac{1}{2}-\Prb{_(}{Z_j,Z_k}{}{-,+}{}{}
}

\noindent
and, from (\ref{eq:BellEspProd})

\eq{}{\nonumber
  \PrbEsp{_(}{Z_j,Z_k,Z_l}{}{Z_j\cdot Z_k}{}{}=-\cos(2\cdot(\theta_k-\theta_j))
}

\noindent
where $\vartheta_j=\theta_j$ and without loss of generality we set $\theta_1=0$, so we get Bell's inequality
(\ref{eq:DesigBell-})

\eq{}{\label{eq:DesigBellFisQExemplo}
  1+\cos(2\cdot\theta_3)\geq\delim{|}{|}{\cos(2\cdot(\theta_2-\theta_3))-\cos(2\cdot\theta_2)}
}

\noindent
and the Wigner's inequalities

\eqn{}{}{\nonumber
  \tfrac{\cos^2(\theta_3)}{2}+\tfrac{\sin^2(\theta_2-\theta_3)}{2}\geq\tfrac{\sin^2(\theta_2)}{2}
  \\\nonumber
  \tfrac{\cos^2(\theta_3-\theta_2)}{2}+\tfrac{\sin^2(\theta_2)}{2}\geq\tfrac{\sin^2(\theta_3)}{2}
}

\begin{figure}
  \centering\includegraphics[scale=0.72]{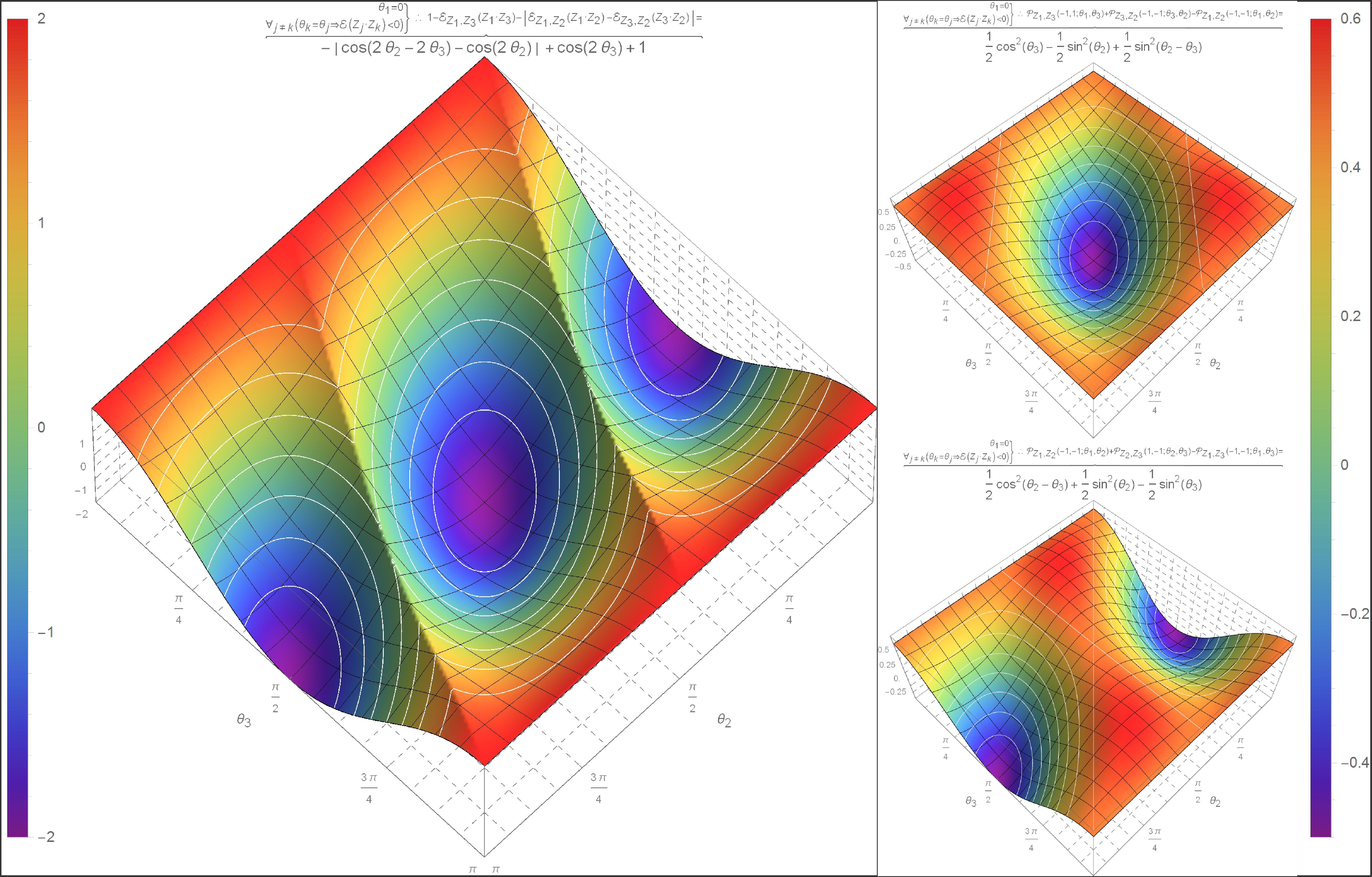}
    \caption{
    The left graph is related to Bell's inequality (\ref{eq:DesigBellFisQExemplo}), in a certain arrangement identified just above the graph. On the right are the graphs of Wigner inequalities with arrangements identified just above the graphs related to the Bell inequality arrangement.
    Bell's region of violation of inequality, region with negative values, coincides with the union of regions of violation, with negative values, of Wigner's inequality in the two arrangements.
    }
  \label{fig:DesigBellWignerFisQ}
\end{figure}

We note that the Bell inequality violation region is the union of equivalent Wigner inequality violation regions.

\subsection{Linear system involving joint and marginal probabilities}\label{sec:Linear system}

One question that may arise is whether it is possible to find values for the probability function, assuming that the marginal probabilities $\Prb{_}{Z_j,Z_k}{}{}{}{}$ are those given by quantum mechanics (\ref{eq:BellProb2upla}). For this, we observe that when assembling the equations relating the joint and marginal probabilities, we have a linear system, given by

\eqn{0pt}{}{\nonumber
  \overbrace{
    \left[\begin{smallmatrix}
      1&1&0&0&0&0&0&0\\
      0&0&0&0&0&0&1&1\\
      1&0&1&0&0&0&0&0\\
      0&0&0&0&0&1&0&1\\
      1&0&0&0&1&0&0&0\\
      0&0&0&1&0&0&0&1\\
      0&0&1&1&0&0&0&0\\
      0&0&0&0&1&1&0&0\\
      0&1&0&1&0&0&0&0\\
      0&0&0&0&1&0&1&0\\
      0&1&0&0&0&1&0&0\\
      0&0&1&0&0&0&1&0
    \end{smallmatrix}\right]
  }^{\mathbf{A}:=}
  \cdot
  \overbrace{
    \left[\begin{smallmatrix}
      \Prb{_(}{Z_1,Z_2,Z_3}{}{-,-,-}{}{}\\
      \Prb{_(}{Z_1,Z_2,Z_3}{}{-,-,+}{}{}\\
      \Prb{_(}{Z_1,Z_2,Z_3}{}{-,+,-}{}{}\\
      \Prb{_(}{Z_1,Z_2,Z_3}{}{-,+,+}{}{}\\
      \Prb{_(}{Z_1,Z_2,Z_3}{}{+,-,-}{}{}\\
      \Prb{_(}{Z_1,Z_2,Z_3}{}{+,-,+}{}{}\\
      \Prb{_(}{Z_1,Z_2,Z_3}{}{+,+,-}{}{}\\
      \Prb{_(}{Z_1,Z_2,Z_3}{}{+,+,+}{}{}
    \end{smallmatrix}\right]
  }^{\mathbf{P}:=}
  =
    \overbrace{
      \left[\begin{smallmatrix}
        \Prb{_(}{Z_1,Z_2}{}{-,-}{}{}\\
        \Prb{_(}{Z_1,Z_2}{}{+,+}{}{}\\
        \Prb{_(}{Z_1,Z_3}{}{-,-}{}{}\\
        \Prb{_(}{Z_1,Z_3}{}{+,+}{}{}\\
        \Prb{_(}{Z_2,Z_3}{}{-,-}{}{}\\
        \Prb{_(}{Z_2,Z_3}{}{+,+}{}{}\\
        \Prb{_(}{Z_1,Z_2}{}{-,+}{}{}\\
        \Prb{_(}{Z_1,Z_2}{}{+,-}{}{}\\
        \Prb{_(}{Z_1,Z_3}{}{-,+}{}{}\\
        \Prb{_(}{Z_1,Z_3}{}{+,-}{}{}\\
        \Prb{_(}{Z_2,Z_3}{}{-,+}{}{}\\
        \Prb{_(}{Z_2,Z_3}{}{+,-}{}{}
      \end{smallmatrix}\right]
    }^{\mathbf{B}}
    =
    \left[\begin{smallmatrix}
      \sin^2(\vartheta_{2}-\vartheta_{1})/2\\ 
      \sin^2(\vartheta_{2}-\vartheta_{1})/2\\ 
      \sin^2(\vartheta_{3}-\vartheta_{1})/2\\ 
      \sin^2(\vartheta_{3}-\vartheta_{1})/2\\ 
      \sin^2(\vartheta_{3}-\vartheta_{2})/2\\ 
      \sin^2(\vartheta_{3}-\vartheta_{2})/2\\ 
      \cos^2(\vartheta_{2}-\vartheta_{1})/2\\ 
      \cos^2(\vartheta_{2}-\vartheta_{1})/2\\ 
      \cos^2(\vartheta_{3}-\vartheta_{1})/2\\ 
      \cos^2(\vartheta_{3}-\vartheta_{1})/2\\ 
      \cos^2(\vartheta_{3}-\vartheta_{2})/2\\ 
      \cos^2(\vartheta_{3}-\vartheta_{2})/2
    \end{smallmatrix}\right]
}

\noindent
where we will omit $(Z_1,Z_2,Z_3)$ in the probabilities of the triples of random variables, writing only $\Prb{(}{}{}{\pm1,\pm1,\pm1}{}{}$, instead of $\Prb{_(}{Z_1,Z_2,Z_3}{}{\pm1,\pm1,\pm1}{}{}$.

From the coefficient matrix we find the following matrix $\mathbf{R}$

\eq{}{\nonumber
  \mathbf{R}:=
  \left[
  \begin{smallmatrix}
    0& 0& 0& 0& 1& 1& 0&-1&-1& 0& 1& 0\\
    0& 0& 0& 0& 0&-1& 0& 0& 1& 0& 0& 0\\
    0& 0& 0& 0& 0&-1& 0& 1& 1&-1&-1& 1\\
    0& 0& 0& 0& 0& 1& 0& 0& 0& 0& 0& 0\\
    0& 0& 0& 0& 0&-1& 0& 1& 1& 0&-1& 0\\
    0& 0& 0& 0& 0& 1& 0& 0&-1& 0& 1& 0\\
    0& 0& 0& 0& 0& 1& 0&-1&-1& 1& 1& 0\\
    1& 0& 0& 0&-1& 0& 0& 1& 0& 0&-1& 0\\
    0& 1& 0& 0& 0&-1& 0& 1& 1&-1&-1& 0\\
    0& 0& 1& 0&-1& 0& 0& 0& 0& 1& 0&-1\\
    0& 0& 0& 1& 0&-1& 0& 0& 1& 0&-1& 0\\
    0& 0& 0& 0& 0& 0& 1&-1&-1& 1& 1&-1
  \end{smallmatrix}
\right]
}

\noindent
which performs the scaling operation when multiplied by the left by it. Thus, we have the scaling of the coefficient matrix $\mathbf{A}$ and the scaling of the augmented matrix $\delim{[}{]}{\tab{}{}{c|c}{\mathbf{A}&\mathbf{B}}}$ by multiplying on the left both by $\mathbf{R}$

\eq{}{\nonumber
  \mathbf{R}\cdot\mathbf{A}=
  \left[
  \begin{smallmatrix}
    1& 0& 0& 0& 0& 0& 0& 1\\
    0& 1& 0& 0& 0& 0& 0&-1\\
    0& 0& 1& 0& 0& 0& 0&-1\\
    0& 0& 0& 1& 0& 0& 0& 1\\
    0& 0& 0& 0& 1& 0& 0&-1\\
    0& 0& 0& 0& 0& 1& 0& 1\\
    0& 0& 0& 0& 0& 0& 1& 1\\
    0& 0& 0& 0& 0& 0& 0& 0\\
    0& 0& 0& 0& 0& 0& 0& 0\\
    0& 0& 0& 0& 0& 0& 0& 0\\
    0& 0& 0& 0& 0& 0& 0& 0\\
    0& 0& 0& 0& 0& 0& 0& 0
  \end{smallmatrix}
  \right]
  ,\quad
  \mathbf{R}\cdot\delim{[}{]}{\tab{}{}{c|c}{\mathbf{A}&\mathbf{B}}}
  =
    \left[\begin{smallmatrix}
      1& 0& 0& 0& 0& 0& 0& 1& 
      (\sin^2(\vartheta_{3}-\vartheta_{1})+\sin^2(\vartheta_{3}-\vartheta_{2})
      -\cos^2(\vartheta_{2}-\vartheta_{1}))/2\\
      0& 1& 0& 0& 0& 0& 0&-1& 
      (1-\sin^2(\vartheta_{3}-\vartheta_{1})-\sin^2(\vartheta_{3}-\vartheta_{2}))/2\\
      0& 0& 1& 0& 0& 0& 0&-1& 
      (\cos^2(\vartheta_{2}-\vartheta_{1})-\sin^2(\vartheta_{3}-\vartheta_{2}))/2\\
      0& 0& 0& 1& 0& 0& 0& 1& \sin^2(\vartheta_{3}-\vartheta_{2})/2\\
      0& 0& 0& 0& 1& 0& 0&-1& (\cos^2(\vartheta_{2}-\vartheta_{1})-\sin^2(\vartheta_{3}-\vartheta_{1}))/2\\
      0& 0& 0& 0& 0& 1& 0& 1& \sin^2(\vartheta_{3}-\vartheta_{1})/2\\
      0& 0& 0& 0& 0& 0& 1& 1& \sin^2(\vartheta_{2}-\vartheta_{1})/2\\
      0& 0& 0& 0& 0& 0& 0& 0& 0\\
      0& 0& 0& 0& 0& 0& 0& 0& 0\\
      0& 0& 0& 0& 0& 0& 0& 0& 0\\
      0& 0& 0& 0& 0& 0& 0& 0& 0\\
      0& 0& 0& 0& 0& 0& 0& 0& 0
    \end{smallmatrix}\right]
}

\noindent
since both matrices have rank 7, then the system can be solved. With rank 7 and 8 probabilities to be defined, we have an undetermined system, where one of the probabilities will be undetermined. So we can write all the other probabilities as a function of this.

\eqn{}{}{\nonumber
  \underbrace{
    \delim{[}{]}{\begin{smallmatrix}
      \Prb{_(}{Z_1,Z_2,Z_3}{}{-,-,-}{}{}+\Prb{_(}{Z_1,Z_2,Z_3}{}{+,+,+}{}{}\\
      \Prb{_(}{Z_1,Z_2,Z_3}{}{-,-,+}{}{}-\Prb{_(}{Z_1,Z_2,Z_3}{}{+,+,+}{}{}\\
      \Prb{_(}{Z_1,Z_2,Z_3}{}{-,+,-}{}{}-\Prb{_(}{Z_1,Z_2,Z_3}{}{+,+,+}{}{}\\
      \Prb{_(}{Z_1,Z_2,Z_3}{}{-,+,+}{}{}+\Prb{_(}{Z_1,Z_2,Z_3}{}{+,+,+}{}{}\\
      \Prb{_(}{Z_1,Z_2,Z_3}{}{+,-,-}{}{}-\Prb{_(}{Z_1,Z_2,Z_3}{}{+,+,+}{}{}\\
      \Prb{_(}{Z_1,Z_2,Z_3}{}{+,-,+}{}{}+\Prb{_(}{Z_1,Z_2,Z_3}{}{+,+,+}{}{}\\
      \Prb{_(}{Z_1,Z_2,Z_3}{}{+,+,-}{}{}+\Prb{_(}{Z_1,Z_2,Z_3}{}{+,+,+}{}{}\\
      0\\
      0\\
      0\\
      0\\
      0
    \end{smallmatrix}}
  }_{=\mathbf{R}\cdot\mathbf{A}\cdot\mathbf{P}}
  =
    \underbrace{
      \delim{[}{]}{\begin{smallmatrix}
        -\Prb{_(}{Z_1,Z_2}{}{+,-}{}{}-\Prb{_(}{Z_1,Z_3}{}{-,+}{}{}
        +\Prb{_(}{Z_2,Z_3}{}{-,-}{}{}+\Prb{_(}{Z_2,Z_3}{}{-,+}{}{}+\Prb{_(}{Z_2,Z_3}{}{+,+}{}{}
        \\
        \Prb{_(}{Z_1,Z_3}{}{-,+}{}{}-\Prb{_(}{Z_2,Z_3}{}{+,+}{}{}
        \\
        \Prb{_(}{Z_1,Z_2}{}{+,-}{}{}+\Prb{_(}{Z_1,Z_3}{}{-,+}{}{}-\Prb{_(}{Z_1,Z_3}{}{+,-}{}{}
        -\Prb{_(}{Z_2,Z_3}{}{-,+}{}{}+\Prb{_(}{Z_2,Z_3}{}{+,-}{}{}-\Prb{_(}{Z_2,Z_3}{}{+,+}{}{}
        \\
        \Prb{_(}{Z_2,Z_3}{}{+,+}{}{}
        \\
        \Prb{_(}{Z_1,Z_2}{}{+,-}{}{}+\Prb{_(}{Z_1,Z_3}{}{-,+}{}{}
        -\Prb{_(}{Z_2,Z_3}{}{-,+}{}{}-\Prb{_(}{Z_2,Z_3}{}{+,+}{}{}
        \\
        -\Prb{_(}{Z_1,Z_3}{}{-,+}{}{}+\Prb{_(}{Z_2,Z_3}{}{-,+}{}{}+\Prb{_(}{Z_2,Z_3}{}{+,+}{}{}
        \\
        -\Prb{_(}{Z_1,Z_2}{}{+,-}{}{}-\Prb{_(}{Z_1,Z_3}{}{-,+}{}{}+\Prb{_(}{Z_1,Z_3}{}{+,-}{}{}
        +\Prb{_(}{Z_2,Z_3}{}{-,+}{}{}+\Prb{_(}{Z_2,Z_3}{}{+,+}{}{}
        \\
        0\\
        0\\
        0\\
        0\\
        0
      \end{smallmatrix}}
    }_{=\mathbf{R}\cdot\mathbf{B}}
}

The last five lines of $\mathbf{R}\cdot\mathbf{B}$ are null (just replace $\Prb{_(}{Z_j,Z_k}{}{z_j,z_k}{}{}=\Prb{_(}{Z_j,Z_k,Z_l}{}{z_j,z_k,-1}{}{}+\Prb{_(}{Z_j,Z_k,Z_l}{}{z_j,z_k,+1}{}{}$ or $\Prb{_(}{Z_j,Z_k}{}{z_j,-1}{}{}+\Prb{_(}{Z_j,Z_k}{}{z_j,+1}{}{}=\Prb{_(}{Z_j}{}{z_j}{}{}$ to check). From this equation we get the expressions that determine the joint probabilities

\eqn{}{}{\nonumber
  \delim{[}{]}{\begin{smallmatrix}
    \Prb{_(}{Z_1,Z_2,Z_3}{}{-,-,-}{}{}\\
    \Prb{_(}{Z_1,Z_2,Z_3}{}{-,-,+}{}{}\\
    \Prb{_(}{Z_1,Z_2,Z_3}{}{-,+,-}{}{}\\
    \Prb{_(}{Z_1,Z_2,Z_3}{}{-,+,+}{}{}\\
    \Prb{_(}{Z_1,Z_2,Z_3}{}{+,-,-}{}{}\\
    \Prb{_(}{Z_1,Z_2,Z_3}{}{+,-,+}{}{}\\
    \Prb{_(}{Z_1,Z_2,Z_3}{}{+,+,-}{}{}\\
  \end{smallmatrix}}
  =
    \delim{[}{]}{\begin{smallmatrix}
      -\Prb{_(}{Z_1,Z_2,Z_3}{}{+,+,+}{}{}
        -\Prb{_(}{Z_1,Z_2}{}{+,-}{}{}-\Prb{_(}{Z_1,Z_3}{}{-,+}{}{}
        +\Prb{_(}{Z_2,Z_3}{}{-,-}{}{}+\Prb{_(}{Z_2,Z_3}{}{-,+}{}{}+\Prb{_(}{Z_2,Z_3}{}{+,+}{}{}
      \\
      \Prb{_(}{Z_1,Z_2,Z_3}{}{+,+,+}{}{}
        +\Prb{_(}{Z_1,Z_3}{}{-,+}{}{}-\Prb{_(}{Z_2,Z_3}{}{+,+}{}{}
      \\
      \Prb{_(}{Z_1,Z_2,Z_3}{}{+,+,+}{}{}
        +\Prb{_(}{Z_1,Z_2}{}{+,-}{}{}+\Prb{_(}{Z_1,Z_3}{}{-,+}{}{}-\Prb{_(}{Z_1,Z_3}{}{+,-}{}{}
        -\Prb{_(}{Z_2,Z_3}{}{-,+}{}{}+\Prb{_(}{Z_2,Z_3}{}{+,-}{}{}-\Prb{_(}{Z_2,Z_3}{}{+,+}{}{}
      \\
      -\Prb{_(}{Z_1,Z_2,Z_3}{}{+,+,+}{}{}
        +\Prb{_(}{Z_2,Z_3}{}{+,+}{}{}
      \\
      \Prb{_(}{Z_1,Z_2,Z_3}{}{+,+,+}{}{}
        +\Prb{_(}{Z_1,Z_2}{}{+,-}{}{}+\Prb{_(}{Z_1,Z_3}{}{-,+}{}{}
        -\Prb{_(}{Z_2,Z_3}{}{-,+}{}{}-\Prb{_(}{Z_2,Z_3}{}{+,+}{}{}
      \\
      -\Prb{_(}{Z_1,Z_2,Z_3}{}{+,+,+}{}{}
        -\Prb{_(}{Z_1,Z_3}{}{-,+}{}{}+\Prb{_(}{Z_2,Z_3}{}{-,+}{}{}+\Prb{_(}{Z_2,Z_3}{}{+,+}{}{}
      \\
      -\Prb{_(}{Z_1,Z_2,Z_3}{}{+,+,+}{}{}
        -\Prb{_(}{Z_1,Z_2}{}{+,-}{}{}-\Prb{_(}{Z_1,Z_3}{}{-,+}{}{}+\Prb{_(}{Z_1,Z_3}{}{+,-}{}{}
        +\Prb{_(}{Z_2,Z_3}{}{-,+}{}{}+\Prb{_(}{Z_2,Z_3}{}{+,+}{}{}\\
    \end{smallmatrix}}
}

\noindent
thus the joint probabilities $\Prb{(}{1,2,3}{}{Z_1=z_1,Z_2=z_2,Z_3=z_3}{}{}$ are determined by the value assigned to $\Prb{(}{1,2,3}{}{Z_1=+1,Z_2=+1,Z_3=+1}{}{}$ and the values assigned to the marginal probabilities $\Prb{(}{j,k}{}{Z_j=z_j,Z_k=z_k}{}{}$.

  \begin{figure}
  
    \centering \includegraphics[scale=0.15]{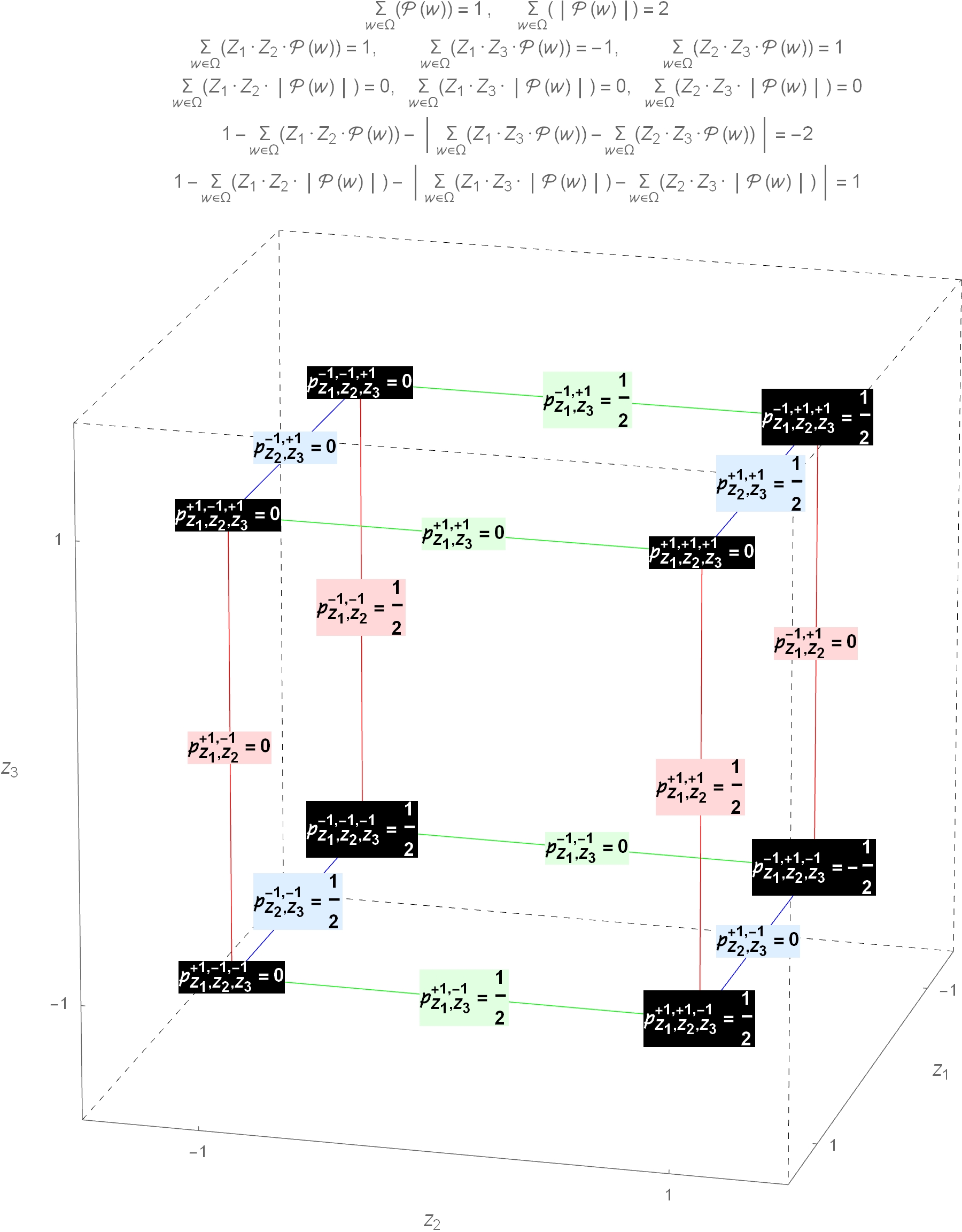}
  \caption[\protect\Prb{}{}{}{}{}{}]{
  The figure shows how quantities are organized: in the cube vertices, in $(z_1,z_2,z_3)\in\{-1,+1\}^3$   are the ``joint odds''
    ($\Prb{(}{}{}{\prbs}{}{}\equiv\Prb{(}{}{}{Z_1(\prbs)=z_1,Z_2(\prbs)=z_2,Z_3(\prbs)=z_3}{}{}\equiv\PrbPhysQ{}{Z_1&,&Z_2&,&Z_3}{}{z_1&,&z_2&,&z_3}{}$)
    and on the colored edges are the marginal odds
    ($\Prb{(}{}{}{Z_j(\prbs)=z_j,Z_k(\prbs)=z_k}{}{}\equiv\PrbPhysQ{}{Z_j&,&Z_k}{}{z_j&,&z_k}{}$).
    The valuation of the ``probabilities'' is obtained according to the marginal
    probability, and just above are the sums of the probabilities (the sum without regard to the absolute
    value 
    $\sum_{\prbs\in\prbS}(\Prb{(}{}{}{\prbs}{}{})$
    the sum that takes into account the absolute value  
    $\sum_{\prbs\in\prbS}(|\Prb{(}{}{}{\prbs}{}{}|)$.
    We also have the correlations, and Bell's inequality, both in the case where we consider such ``probabilities'' (even if negative in this case) and in the case in which we consider their absolute values (\ref{eq:DesigBellGeral}) 
    (where we note that no there is violation of inequality). }
  
  \label{fig:ExemploDesigBellFisQ}
  \end{figure}

In the figure \ref{fig:ExemploDesigBellFisQ}, we have an example of valuation, where we can observe that, according to Bell's inequality (\ref{eq:DesigBellGeral}) (taking care to observe the conditions used in the demonstration (\ref{eq:DesigBellDemo})), the inequality is not violated, however, if substitution is made
without paying attention to the conditions, then the violation of inequality arises.

We then conclude that we can assign values to $\Prb{(}{1,2,3}{}{Z_1=z_1,Z_2=z_2,Z_3=z_3}{}{}$ so that we have $\Prb{(}{j,k}{}{Z_j=z_j,Z_k=z_k}{}{}$, however, for some values of $\theta$, one of Kolmogorov's axioms is violated, ie $\Prb{(}{1,2,3}{}{Z_1=z_1,Z_2=z_2,Z_3=z_3}{}{}\ngeq0$.

\section{Statistical estimators}\label{Sec:Statistical estimators}

In practice, when conducting an experiment, what is used are estimators, so the values assigned to the probabilities are estimated values. In this case, we have the following estimators

\eq{}{\nonumber
  \Prb{_(}{Z_j,Z_k}{}{-,-}{}{}
  \esteq
    \tfrac{n_{j,k}^{--}}{N}
  ,\ 
  \Prb{_(}{Z_j,Z_k}{}{-,+}{}{}
  \esteq
    \tfrac{n_{j,k}^{-+}}{N}
  ,\ 
  \Prb{_(}{Z_j,Z_k}{}{+,-}{}{}
  \esteq
    \tfrac{n_{j,k}^{+-}}{N}
  ,\ 
  \Prb{_(}{Z_j,Z_k}{}{+,+}{}{}
  \esteq
    \tfrac{n_{j,k}^{++}}{N}
  ,\\\nonumber
  \PrbEsp{(}{}{}{Z_j\cdot Z_k}{}{}
  \esteq
    \tfrac{n_{j,k}^{--}-n_{j,k}^{-+}-n_{j,k}^{+-}+n_{j,k}^{++}}{N}
}

\noindent
where: $n_{j,k}^{-,-}$ is the number of occurrences of the event where $(Z_j,Z_k)=(-1,-1)$, $n_{j,k}^{-,+}$ is the number of occurrences of the event where $(Z_j,Z_k)=(-1,+1)$, $n_{j,k}^{+,-}$ is the number of occurrences of the event. event where $(Z_j,Z_k)=(+1,-1)$, $n_{j,k}^{+,+}$ is the number of occurrences of the event where $(Z_j,Z_k)=(+1,+1)$ and $N$ is the number of times the experiment was performed.

\eq{}{\nonumber
  N
  =
    n_{1,3}^{--}+n_{1,3}^{-+}+n_{1,3}^{+-}+n_{1,3}^{++}
    +n_{1,2}^{--}+n_{1,2}^{-+}+n_{1,2}^{+-}+n_{1,2}^{++}
    +n_{2,3}^{--}+n_{2,3}^{-+}+n_{2,3}^{+-}+n_{2,3}^{++}
}

By replacing the estimators on Bell's inequality, we have

\eqn{}{}{\nonumber
  &&\overbrace{
    1-\tfrac{n_{1,3}^{--}-n_{1,3}^{-+}-n_{1,3}^{+-}+n_{1,3}^{++}}{N}
  }^{1-\PrbEsp{(}{}{}{Z_1\cdot Z_3}{}{}\esteq}
  \geq
    \overbrace{
      \delim{|}{|}{\tfrac{n_{1,2}^{--}-n_{1,2}^{-+}-n_{1,2}^{+-}+n_{1,2}^{++}}{N}
      -\tfrac{n_{2,3}^{--}-n_{2,3}^{-+}-n_{2,3}^{+-}+n_{2,3}^{++}}{N}}
    }^{|\PrbEsp{(}{}{}{Z_1\cdot Z_2}{}{}-\PrbEsp{(}{}{}{Z_2\cdot Z_3}{}{}|\esteq}
  \If\\\nonumber
  &&\If
    \delim{\{}{}{\tab{0pt}{}{l}{
      1
      \geq
        \tfrac{n_{1,3}^{--}-n_{1,3}^{-+}-n_{1,3}^{+-}+n_{1,3}^{++}}{N}
        +\delim{(}{)}{\frac{n_{1,2}^{--}-n_{1,2}^{-+}-n_{1,2}^{+-}+n_{1,2}^{++}}{N}
        -\tfrac{n_{2,3}^{--}-n_{2,3}^{-+}-n_{2,3}^{+-}+n_{2,3}^{++}}{N}}
      \\
      1
      \geq
        \tfrac{n_{1,3}^{--}-n_{1,3}^{-+}-n_{1,3}^{+-}+n_{1,3}^{++}}{N}
        -\delim{(}{)}{\frac{n_{1,2}^{--}-n_{1,2}^{-+}-n_{1,2}^{+-}+n_{1,2}^{++}}{N}
        -\tfrac{n_{2,3}^{--}-n_{2,3}^{-+}-n_{2,3}^{+-}+n_{2,3}^{++}}{N}}
    }}
}

\eqn{}{}{\label{eq:BellDesigualdadeEstimadores1}
  N
  \geq
    n_{1,3}^{--}-n_{1,3}^{-+}-n_{1,3}^{+-}+n_{1,3}^{++}
    +n_{1,2}^{--}-n_{1,2}^{-+}-n_{1,2}^{+-}+n_{1,2}^{++}
    -n_{2,3}^{--}+n_{2,3}^{-+}+n_{2,3}^{+-}-n_{2,3}^{++}
  \\\nonumber
  N
  \geq
    n_{1,3}^{--}-n_{1,3}^{-+}-n_{1,3}^{+-}+n_{1,3}^{++}
    -n_{1,2}^{--}+n_{1,2}^{-+}+n_{1,2}^{+-}-n_{1,2}^{++}
    +n_{2,3}^{--}-n_{2,3}^{-+}-n_{2,3}^{+-}+n_{2,3}^{++}
}

\noindent
therefore, being $N$ and the $n$'s all natural (clearly nonnegative) numbers, it would not be possible for inequality to be experimentally violated unless $N$ is not the total number of experiments.

In the quantum mechanics literature, we find the use of $ n_ {j, k} $ (number of experiments in one of the configurations)

\eq{}{\nonumber
  n_{j,k}
  =
    n_{j,k}^{--}+n_{j,k}^{-+}+n_{j,k}^{+-}+n_{j,k}^{++}
    ,\qquad
    (j,k)\in\{(1,2),(1,3),(2,3)\}
}

\noindent
instead of $N$, therefore, for a given pair $(j,k)$, the number of times the experiment was performed in the setting related to that pair is counted, ignoring the other settings. However, when $n_{1,2}\neq n_{1,3}\neq n_{2,3}$, the question arises: which one would replace $N$? Because we would have 
$\PrbEsp{(}{}{}{Z_1\cdot Z_2-Z_2\cdot Z_3}{}{}
 =
  \tfrac
  {n_{1,2}^{--}-n_{1,2}^{-+}-n_{1,2}^{+-}+n_{1,2}^{++}}{n_{1,2}}
  -\tfrac{n_{2,3}^{--}-n_{2,3}^{-+}-n_{2,3}^{+-}+n_{2,3}^{++}}{n_{1,2}}$
or 
$\PrbEsp{(}{}{}{Z_1\cdot Z_2-Z_2\cdot Z_3}{}{}
 =
  \tfrac
  {n_{1,2}^{--}-n_{1,2}^{-+}-n_{1,2}^{+-}+n_{1,2}^{++}}{n_{2,3}}-\tfrac{n_{2,3}^{--}-n_{2,3}^{-+}-n_{2,3}^{+-}+n_{2,3}^{++})}{n_{2,3}}$,
which would differ from each other in this case. Or maybe it would be 
$\PrbEsp{(}{}{}{Z_1\cdot Z_2-Z_2\cdot Z_3}{}{}
 =
  \tfrac
  {(n_{1,2}^{--}-n_{1,2}^{-+}-n_{1,2}^{+-}+n_{1,2}^{++})-(n_{2,3}^{--}-n_{2,3}^{-+}-n_{2,3}^{+-}+n_{2,3}^{++})}
  {n_{1,2}+n_{2,3}}
$,
$\PrbEsp{(}{}{}{Z_1\cdot Z_2}{}{}
 =
  \tfrac
  {n_{1,2}^{--}-n_{1,2}^{-+}-n_{1,2}^{+-}+n_{1,2}^{++}}
  {n_{1,2}}$
and 
$\PrbEsp{(}{}{}{Z_2\cdot Z_3}{}{}
 =
  \tfrac
  {n_{2,3}^{--}-n_{2,3}^{-+}-n_{2,3}^{+-}+n_{2,3}^{++}}
  {n_{2,3}}$
making $\PrbEsp{(}{}{}{Z_1\cdot Z_2-Z_2\cdot Z_3}{}{}$ different from $\PrbEsp{(}{}{}{Z_1\cdot Z_2}{}{}-\PrbEsp{(}{}{}{Z_2\cdot Z_3}{}{}$.

We often find experiments where $n_{1,2}=n_{1,3}=n_{2,3}=n$. Statistically, replacing $N$ with $n_{j,k}$ makes no sense unless we are dealing with conditional expected values. We will see below what this implies.

The proportion of times the experiment was performed in the $(j,k)$ pair-related configuration is given by $\frac{n_{j,k}}{N}$, which can be interpreted as the probability of experiment is in the $(j,k)$ pair-related configuration (even though such a configuration is not random, although in some cases it is). In the case where the experiment setup is not random, we can still interpret it as a probability, just as we use conditional probability to refer to events that have already occurred (so they are events with probability 1). So we have the following probabilities

\eq{}{\nonumber
  \tab{}{}{lll}{
    \Prb{(}{}{}{j=1,k=2}{}{}
    \esteq
      \tfrac{n_{1,2}}{N}
    ,\quad
    &\Prb{(}{}{}{j=1,k=3}{}{}
    \esteq
      \tfrac{n_{1,3}}{N}
    ,\quad
    &\Prb{(}{}{}{j=2,k=3}{}{}
    \esteq
      \tfrac{n_{2,3}}{N}
  }
}

\noindent
where $\tfrac{n_{j,k}}{N}$ is an estimate of the probability of finding the experiment in the setting related to the pair $(j,k)$. We also have the conditional probabilities

\eq{}{\nonumber
  \tab{2pt}{}{lll}{
    \Prb{;(;}{}{}{Z_j=a,Z_k=b}{j=1,k=2}{}
    \esteq
      \tfrac{n_{1,2}^{a,b}}{n_{1,2}}
    ,
    &\Prb{;(;}{}{}{Z_j=a,Z_k=b}{j=1,k=3}{}
    \esteq
      \tfrac{n_{1,3}^{a,b}}{n_{1,3}}
    ,
    &\Prb{;(;}{}{}{Z_j=a,Z_k=b}{j=2,k=3}{}
    \esteq
      \tfrac{n_{2,3}^{a,b}}{n_{2,3}}
  }
}

\noindent
where $\tfrac{n_{j,k}^{a,b}}{n_{j,k}}$ is the conditional probability (probability of occurring $(Z_j,Z_k)=(a,b)$ since we are in an experiment related to $(j,k)$, among the three possible configurations: $(j,k)$, $(j,l)$, and $(k,l)$).

\subsection{Loopholes}\label{Loopholes}

As we can see below, when we consider the additional possibility of $Z_j=0$, we have

\eqn{0pt}{}{\nonumber
  \delim{}{\}}{\tab{}{}{lll}{
    Z_1=0
    \If
    1\pm\underbrace{Z_1\cdot Z_3}_{=0}
    \geq
      |\underbrace{Z_1\cdot Z_2}_{=0}\pm Z_2\cdot Z_3|
    &\If&
    1\geq|Z_2\cdot Z_3|
    \\
    Z_2=0
    \If
    1\pm Z_1\cdot Z_3
    \geq
      |\underbrace{Z_1\cdot Z_2}_{=0}\pm\underbrace{Z_2\cdot Z_3}_{=0}|
    &\If&
    \underbrace{1\geq\mp Z_1\cdot Z_3}_{1\geq|Z_1\cdot Z_3|}
    \\
    Z_3=0
    \If
    1\pm\underbrace{Z_1\cdot Z_3}_{=0}
    \geq
      |Z_1\cdot Z_2\pm\underbrace{Z_2\cdot Z_3}_{=0}|
    &\If&
    1\geq|Z_1\cdot Z_2|
  }}
  \therefore
  Z_l=0\If 1\geq|Z_j\cdot Z_k|
}

\noindent
thus, in all cases $Z_l=0$ result in inequalities $1\geq|Z_j\cdot Z_k|$ which are valid because $\fa{j}{Z_j\in\{-1,0,+1\}}$, consequently, Bell's inequality will also be valid. So now we can include cases where there is only one detection $(Z_j,Z_k,Z_l)=(\pm1,0,0)$.

\eqn{}{}{\nonumber
  Z_l=0
  \If 1\geq\overbrace{|Z_j\cdot Z_k|=0}^{(Z_l=0\Or Z_k=0)\If}
}

So if in addition to $Z_l=0$ we have $Z_l=0 $ or $ Z_k = 0 $ or both, the inequality is still valid, resulting in $ 1 \geq 0 $. Therefore, it is concluded that

\eqn{}{}{\nonumber
    (Z_j,Z_k,Z_l)\in\{-1,+1\}^3
  &\If&
  1\pm Z_j\cdot Z_l\equiv|Z_j\cdot Z_k\pm Z_k\cdot Z_l|
  \\\nonumber
  (Z_j,Z_k,Z_l)\in\{-1,0,+1\}^3
  &\If&
  1\pm Z_j\cdot Z_l\geq|Z_j\cdot Z_k\pm Z_k\cdot Z_l|
}

\noindent
for $(Z_1,Z_2,Z_3)\in\{-1,0,+1\}^3$, so we can model cases where there is no detection in one or more detectors. Such modeling allows us to assign probability to these cases, where we can interpret as cases where the efficiency of the detectors is not $100 \%$, and to obtain Bell's inequality, we proceed in the same way as in (\ref{eq:DesigBellDemo}) , where we find that

\eqn{}{}{\nonumber
  1\pm\PrbEsp{(}{}{}{Z_j\cdot Z_l}{}{}
  \geq
    |\PrbEsp{(}{}{}{Z_j\cdot Z_k}{}{}\pm\PrbEsp{(}{}{}{Z_k\cdot Z_l}{}{}|
}

\noindent
whatever the probability function (including extreme cases where probability is assigned to event $(Z_1,Z_2,Z_3)\in\{-1,+1\}^3$ of having 3 detections and event $ (Z_1, Z_2, Z_3) = (0,0,0) $ of not having any), just that it is defined for random variables $Z_j:\prbS\rightarrow\{-1,0,+1\}$. Thus, such reshaping of the random variable allows more interpretations and allows to expand the inequality of Bell for situations that were not being considered.

The expected value $\PrbEsp{(}{}{}{Z_j\cdot Z_k}{}{}$ will be given by

\eqn{0pt}{}{\nonumber
  \PrbEsp{(}{}{}{Z_j\cdot Z_k}{}{}
  &=&
    \sum\limits_{\tab{0pt}{0}{c}{\scriptstyle z_j\in\{-1,0,+1\}\\\scriptstyle z_k\in\{-1,0,+1\}}}\Bigg(
      z_j\cdot z_k\cdot
      \sum\limits_{z_l\in\{-1,0,+1\}}\delim{(}{)}{
        \Prb{_(}{Z_j,Z_k,Z_l}{}{z_j,z_k,z_l}{}{}
      }
    \Bigg)
  \\\nonumber
  &=&
    \sum\limits_{\tab{0pt}{0}{c}{\scriptstyle z_j\in\{-1,+1\}\\\scriptstyle z_k\in\{-1,+1\}}}\Bigg(
      z_j\cdot z_k\cdot
      \underbrace{
        \sum\limits_{z_l\in\{-1,0,+1\}}\delim{(}{)}{
          \Prb{_(}{Z_j,Z_k,Z_l}{}{z_j,z_k,z_l}{}{}
        }
      }_{=\Prb{(}{Z_j,Z_k}{}{Z_j=z_j,Z_k=z_k}{}{}}
    \Bigg)
  \\\nonumber
  &=&
    \Prb{_(}{Z_j,Z_k}{}{-,-}{}{}+\Prb{_(}{Z_j,Z_k}{}{+,+}{}{}
    -\Prb{_(}{Z_j,Z_k}{}{+,-}{}{}-\Prb{_(}{Z_j,Z_k}{}{-,+}{}{}
}

\noindent
which apparently is the same expression that we had, once

\eqn{0pt}{}{\nonumber
  (z_j=0\Or z_k=0)
  \If \underbrace{z_j\cdot z_k}_{=0}\cdot\Prb{(}{}{}{Z_j=z_j,Z_k=z_k}{}{}=0
}

\noindent
therefore the probabilities that one of the random variables cancel each other out does not appear in the expression,
however, the $\Prb{(}{}{}{Z_j=z_j,Z_k=z_k}{}{}$ (where $(z_j,z_k)\in\{-1,+1\}$) probabilities do not have the same values, as this modeling takes into account cases where only one detection occurs (when detector efficiency is not $100 \%$), and other cases, such as There is detection in all 3 detectors (in the event of a failure or event outside the experiment interfering).

Considering the statistical estimators, it is more evident the decrease that the $\Prb{(}{}{}{Z_j=z_j,Z_k=z_k}{}{}$ probabilities suffer due to the drop in detector efficiency. Initially, let's define the following quantity: $n_{j,k}^{\alpha_j\alpha_k}$ is the number of $(Z_j,Z_k)=(z_j,z_k)$-related event occurrences, where $(z_j=-1\If\alpha_j=-)$, $(z_j=0\If\alpha_j=0)$ and $(z_j=+1\If\alpha_j=+)$, and analogously for $\alpha_k$.
So we have the following estimators

\eqn{}{}{\nonumber
  \tab{4pt}{}{*{6}{r}}{
    \Prb{_(}{Z_j,Z_k}{}{-1,-1}{}{}=\frac{n_{j,k}^{--}}{n_{j,k}},
    &\Prb{_(}{Z_j,Z_k}{}{-1,0}{}{}=\frac{n_{j,k}^{-0}}{n_{j,k}},
    &\Prb{_(}{Z_j,Z_k}{}{-1,+1}{}{}=\frac{n_{j,k}^{-+}}{n_{j,k}},
    \\
    \Prb{_(}{Z_j,Z_k}{}{0,-1}{}{}=\frac{n_{j,k}^{0-}}{n_{j,k}},
    &\Prb{_(}{Z_j,Z_k}{}{0,0}{}{}=\frac{n_{j,k}^{00}}{n_{j,k}},
    &\Prb{_(}{Z_j,Z_k}{}{0,+1}{}{}=\frac{n_{j,k}^{0+}}{n_{j,k}},
    \\
    \Prb{_(}{Z_j,Z_k}{}{+1,-1}{}{}=\frac{n_{j,k}^{+-}}{n_{j,k}},
    &\Prb{_(}{Z_j,Z_k}{}{+1,0}{}{}=\frac{n_{j,k}^{+0}}{n_{j,k}},
    &\Prb{_(}{Z_j,Z_k}{}{+1,+1}{}{}=\frac{n_{j,k}^{++}}{n_{j,k}}
  }
  \\\nonumber
  n_{j,k}=n_{j,k}^{--}+n_{j,k}^{-0}+n_{j,k}^{-+}+n_{j,k}^{0-}+n_{j,k}^{00}+n_{j,k}^{0+}+n_{j,k}^{+-}+n_{j,k}^{+0}+n_{j,k}^{++}
  \\\nonumber
  n_{j,k}^{\alpha_j\alpha_k}=n_{j,k,l}^{\alpha_j\alpha_k-}+n_{j,k,l}^{\alpha_j\alpha_k0}+n_{j,k,l}^{\alpha_j\alpha_k+}
}

We note that the estimators are similar to what was presented, however $ n_{j, k} $ is now the sum of some events that were not previously considered, which are those that are not detected in one or more detectors.

When assigning the values belonging to $\delim{\{}{\}}{k/10}_{k=0}^{k=10}$ to the 26 probabilities $\Prb{(}{}{}{Z_1=z_1,Z_2=z_2,Z_3=z_3}{}{}$ (the probability $\Prb{(}{}{}{Z_1=0,Z_2=0,Z_3=0}{}{}$ is defined after determining the others, in such a way that the sum of all probabilities results in 1), we can observe that, whatever the values assigned, in none of the $\frac{(10+26)!}{10!\cdot26!}=254186856$ possible combinations there was a violation of Bell's inequality (that is, occurrence of negative values).

\begin{figure}
  \centering\includegraphics[scale=0.15]{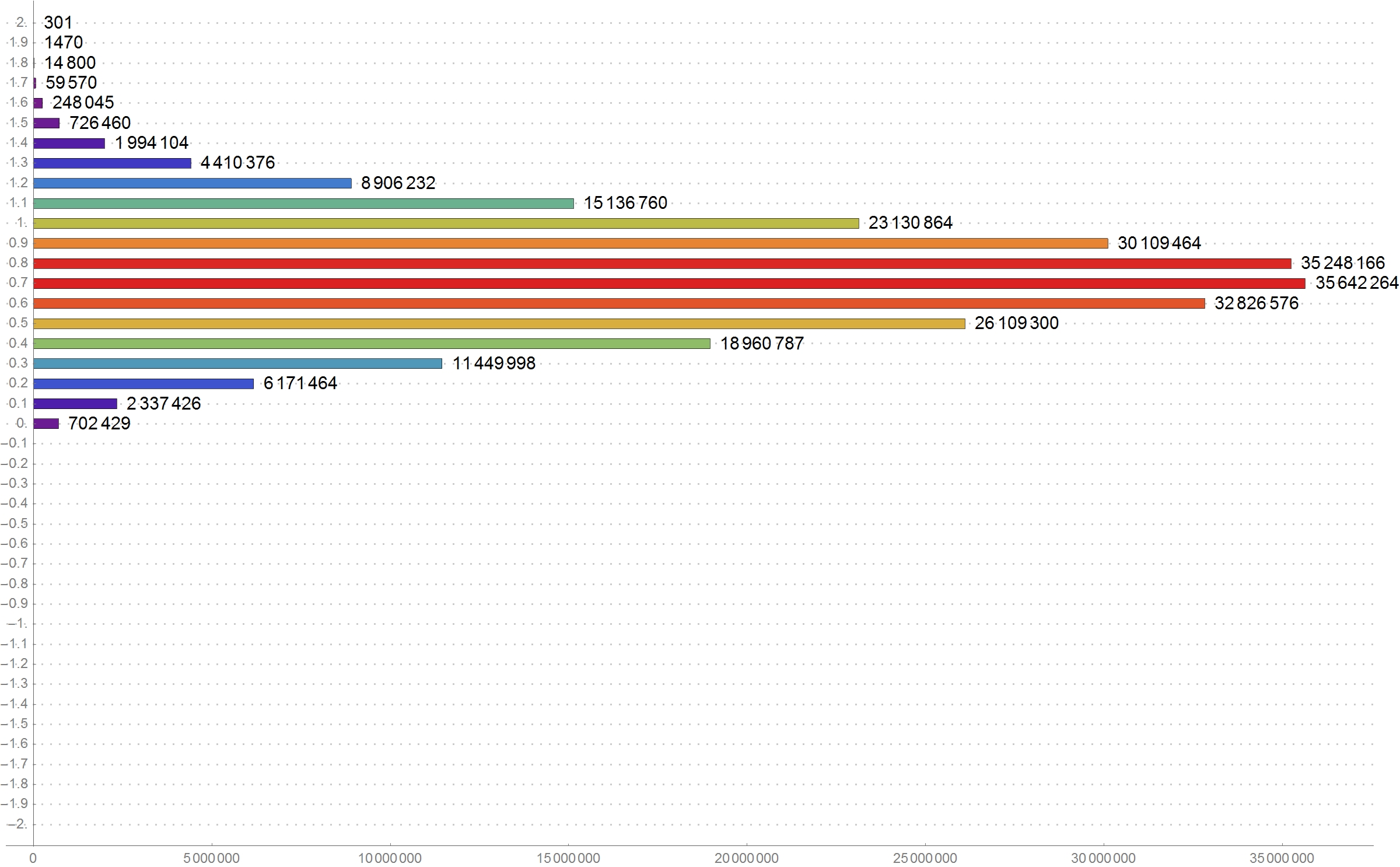}
  \caption[\protect\PrbEsp{}{}{}{}{}{}]{
    Histogram of the frequency of the values assumed by the first member of Bell's inequality $1-\PrbEsp{(}{}{}{Z_1\cdot Z_3}{}{}-|\PrbEsp{(}{}{}{Z_1\cdot Z_2}{}{}-\PrbEsp{(}{}{}{Z_2\cdot Z_3}{}{}|\geq0,$ performing all combinations of value assignments $\delim{\{}{\}}{k/10}_{k=0}^{k=10}$ to the probabilities  $\Prb{(}{}{}{Z_1=z_1,Z_2=z_2,Z_3=z_3}{}{}$. On the vertical axis are the classes (range of Bell inequality values) and on the horizontal axis are the absolute frequencies. There were no negative values.
  }
\end{figure}

\subsection{Conditional probability}\label{Sec:Conditional probability}

In this section, through the estimates presented in the previous section, we propose that the probabilities involved in violating Bell's inequality should be interpreted as conditional probabilities, so the expected values calculated based on them are conditional expected values, which will be used to find expected values of Bell's inequality. From the conditional probability $\Prb{;(;}{}{}{(Z_1,Z_2,Z_3)\in A}{(Z_1,Z_2,Z_3)\in B}{}$ which is defined by

\eq{}{\nonumber
  \Prb{;(;}{}{}{(Z_1,Z_2,Z_3)\in A}{(Z_1,Z_2,Z_3)\in B}{}
  \defeq
    \frac{\Prb{(}{}{}{(Z_1,Z_2,Z_3)\in A\cap B}{}{}}{\Prb{(}{}{}{(Z_1,Z_2,Z_3)\in B}{}{}}
  ,\quad
  \Prb{(}{}{}{(Z_1,Z_2,Z_3)\in B}{}{}
  \neq
    0
}

\noindent
that in our case $B$ will represent the configuration in which the experiment was performed, ie $(j,k)=(1,2)$ represents that the apparatus detectors $j=1$ and $k=2$ detected both particles ($(Z_1,Z_2)\in\{-1,+1\}^2$), consequently, apparatus 3 detector did not detect any particles ($Z_3=0$). So $\Prb{;(;}{}{}{Z_j=z_j,Z_k=z_k}{j,k}{} \equiv\Prb{;(;}{}{}{Z_j=z_j,Z_k=z_k}{Z_l=0}{}$ and $\Prb{(}{}{}{j,k}{}{} \equiv\Prb{(}{}{}{Z_l=0}{}{}$ where $\{j,k,l\}=\{1,2,3\}$.

This results in a modification of the random variables because previously we had considered that $Z_j\in\{-1,+1\}$ (with $j\in\{1,2,3\}$), but now we consider that $Z_j\in\{-1,0,+1\}$, where $Z_j=0$ will mean no particle has interacted with the $j$ apparatus, and all other cases continue to be interpreted as before. Such a modification is compatible with Bell's inequality. 

By the definition of conditional probability, we have

\eqn{}{}{\nonumber
  \Prb{;(;}{}{}{Z_j=z_j,Z_k=z_k}{Z_l=0}{}
  =
    \tfrac{\Prb{(}{}{}{Z_j=z_j,Z_k=z_k,Z_l=0}{}{}}{\Prb{(}{}{}{Z_l=0}{}{}}
  \If\\\nonumber
  \If
  \Prb{(}{}{}{Z_j=z_j,Z_k=z_k,Z_l=0}{}{}
  =
    \Prb{;(;}{}{}{Z_j=z_j,Z_k=z_k}{Z_l=0}{}\cdot\Prb{(}{}{}{Z_l=0}{}{}
}

\noindent
where the joint probability $\Prb{(}{}{}{Z_j=z_j,Z_k=z_k,Z_l=0}{}{}$ represents the probability that no detection on the $l$ apparatus will occur, or no, some detection in the $j$ and $k$ apparatus, depending on the value assumed by $(z_j,z_k)$. The conditional probability $\Prb{;(;}{}{}{Z_j=z_j,Z_k=z_k}{Z_l=0}{}$ represents the probability of occurrence, or not, of the some detection in the $j$ and $k$ apparatus, disregarding cases where there is some detection in the $l$ apparatus (ie $Z_l\neq0$). Already the marginal probability $\Prb{(}{}{}{Z_l=0}{}{} \equiv\sum_{(z_j,z_k)\in\{-1,0,+1\}}\delim{(}{)}{\Prb{(}{}{}{Z_j=z_j,Z_k=z_k,Z_l=0}{}{}}$ represents the probability that there is no detection on the $l$ apparatus, considering all possibilities in relation to other apparatuses.

Since we have the conditional probabilities (where the experiment is kept in a given configuration) and the marginal probabilities (the probabilities, or the proportions of times the experiment is in each of the configurations), we can find the combined probabilities of $(Z_j,Z_k,Z_l)$ using full probability law

\eqn{0pt}{}{\label{eq:ProbCond}
  \Prb{(}{}{}{Z_j=z_j,Z_k=z_k,Z_l=z_l}{}{}
  &=&
    \sum\limits_{z\in\{-1,0,+1\}}\delim{}{}{
      \Prb{;(;}{}{}{Z_j=z_j,Z_k=z_k,Z_l=z_l}{Z_l=z}{}\cdot\Prb{(}{}{}{Z_l=z}{}{}
    }
  \\\nonumber
  &=&
    \delim{\{}{}{\tab{}{}{ll}{
      \Prb{;(;}{}{}{Z_j=z_j,Z_k=z_k}{Z_l=0}{}\cdot\Prb{(}{}{}{Z_l=0}{}{}
      &\fI z_l=z=0
      \\
      \Prb{;(;}{}{}{Z_j=z_j,Z_k=z_k}{Z_l=-1}{}\cdot\Prb{(}{}{}{Z_l=-1}{}{}
      &\fI z_l=z=-1
      \\
      \Prb{;(;}{}{}{Z_j=z_j,Z_k=z_k}{Z_l=+1}{}\cdot\Prb{(}{}{}{Z_l=+1}{}{}
      &\fI z_l=z=+1
    }}
}

Now, considering that there is never detection in all three apparatuses simultaneously, we have

\eq{}{\nonumber
  \Prb{(}{}{}{Z_j=z_j,Z_k=z_k,Z_l=-1}{}{}=\Prb{(}{}{}{Z_j=z_j,Z_k=z_k,Z_l=+1}{}{}=0
  ,\quad(z_j,z_k)\in\{-1,+1\}^2
}

\noindent
therefore

\eq{}{\nonumber
  (z_j,z_k)\in\{-1,+1\}
  \If
  \Prb{(}{Z_j,Z_k,Z_l}{}{Z_j=z_j,Z_k=z_k,Z_l=z_l}{}{}
  =
    \delim{\{}{}{\tab{0pt}{}{ll}{
      \Prb{;(;}{}{}{Z_j=z_j,Z_k=z_k}{Z_l=0}{}\cdot\Prb{(}{}{}{Z_l=0}{}{}
      &\fI z_l=0
      \\
      0
      &\fI z_l\neq0
    }}
}

  \begin{figure}
  
    \centering\includegraphics[scale=0.15]{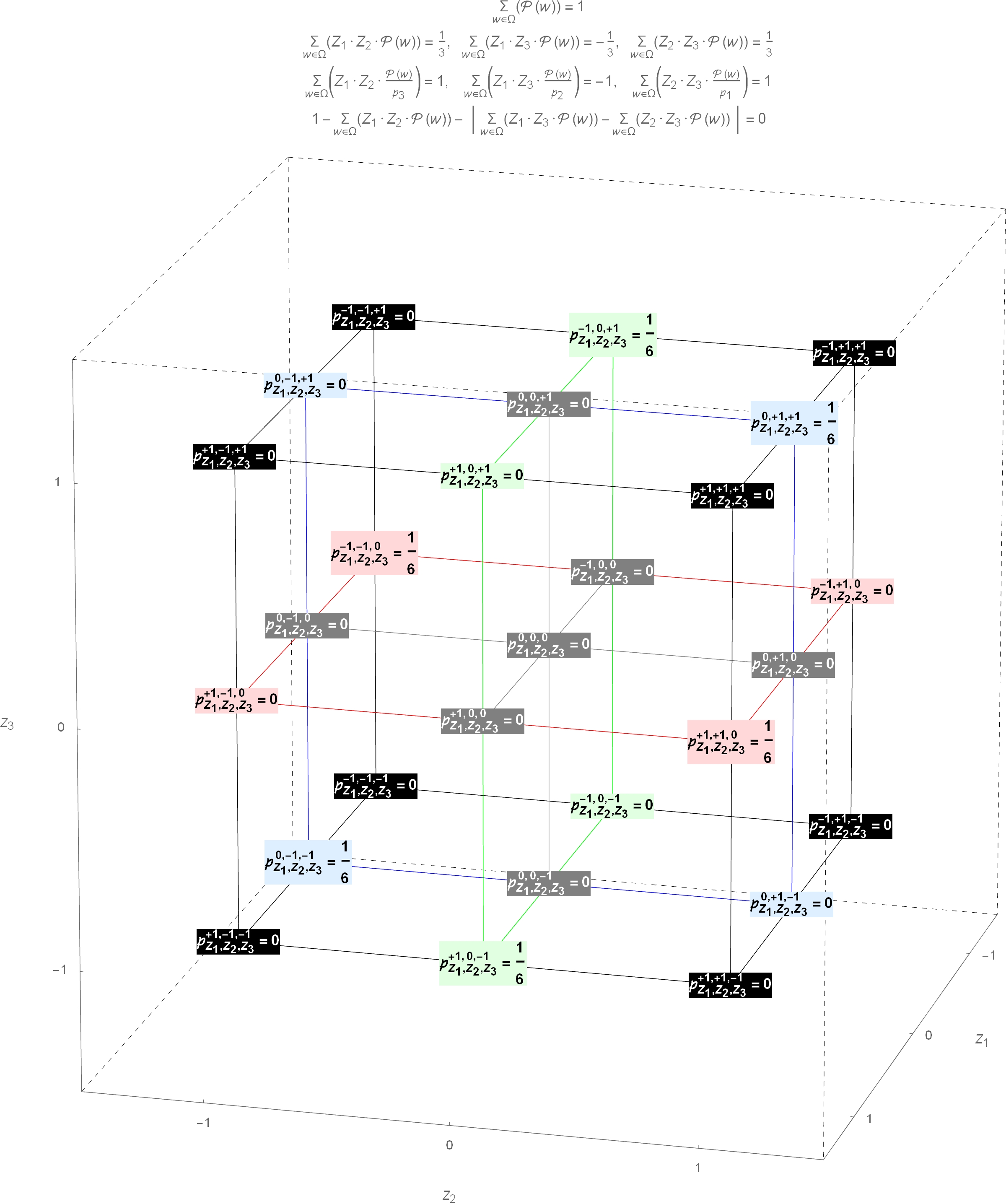}
      
  \caption{
  At each point $(z_1,z_2,z_3)\in\{-1,0,+1\}^3$ we have an equation where the left member shows how the quantities are arranged, we use the same notation adopted in the previous figure.
     On the right member is a case of probability valuation, which was obtained according to the same marginal probabilities used in the previous figure, and just above are the sums of probabilities.
     We also have the correlations, and Bell's inequality.
  }
  
  \end{figure}

The expected value of $Z_j\cdot Z_k$ will be

\eqn{}{}{\nonumber
  \PrbEsp{(}{}{}{Z_j\cdot Z_k}{}{}
  =
    \sum\limits_{(z_j,z_k,z_l)\in\{-1,0,+1\}^3}\delim{(}{)}{
      z_j\cdot z_k\cdot\Prb{(}{}{}{Z_j=z_j,Z_k=z_k,Z_l=z_l}{}{}
    }
}

\noindent
when $z_j=0$ or $z_k=0$, the product $z_j\cdot z_k\cdot\Prb{}{}{}{}{}{}$ is null, so we only consider cases $(z_j,z_k)\in\{-1,+1\}$, which will result in

\eqn{}{}{\nonumber
  \PrbEsp{(}{}{}{Z_j\cdot Z_k}{}{}
  =
    \underbrace{
      \sum\limits_{(z_j,z_k)\in\{-1,+1\}^2}\delim{(}{)}{
        z_j\cdot z_k\cdot\Prb{;(;}{}{}{Z_j=z_j,Z_k=z_k}{Z_l=0}{}{}
      }
    }_{=\PrbEsp{;(;}{}{}{Z_j\cdot Z_k}{Z_l=0}{}}
    \cdot\Prb{(}{}{}{Z_l=0}{}{}
}

\noindent
where $\PrbEsp{;(;}{}{}{Z_j\cdot Z_k}{Z_l=0}{}$ is the conditional expected value, which is nothing more than the expected value calculated based on a conditional probability.

Using the usual statistical notation\footnote{
  The index indicates the random variables in which the probability function is defined and the argument is the expression of the random variables in which we calculate the expected value: $\PrbEsp{_(}{Z_j,Z_k}{}{f(Z_j,Z_k)}{}{}= \sum_{(z_j,z_k)\in\{-1,0,+1\}^2}\delim{(}{)}{f(z_j,z_k)\cdot\Prb{(}{}{}{Z_j=z_j,Z_k=z_k}{}{}}$ For the conditional expected value, we have the following notation $\PrbEsp{;_(;}{Z_j}{Z_k}{f(Z_j,Z_k)}{Z_k=z_k}{}= \sum_{z_j\in\{-1,0,+1\}}\delim{(}{)}{f(z_j,z_k)\cdot\tfrac{\Prb{(}{}{}{Z_j=z_j,Z_k=z_k}{}{}}{\Prb{(}{}{}{Z_k=z_k)}{}{}}}$ where $z_k$ is fixed.
}, we have the following formula, which lists the expected value $\PrbEsp{_(}{Z_j,Z_k}{}{Z_j\cdot Z_k}{}{}$ (calculated from the joint probability $\Prb{_(}{Z_j,Z_k,Z_l}{}{z_j,z_k,z_l}{}{}$) with the expected conditional value $\PrbEsp{;_(;}{Z_j,Z_k}{Z_l}{Z_j\cdot Z_k}{0}{}$

\eqn{0pt}{}{\label{eq:Valor esperado e valor esperado condicional}
  \PrbEsp{_(}{Z_j,Z_k,Z_l}{}{Z_j\cdot Z_k}{}{}
  &\equiv&
    \PrbEsp{_(}{Z_l}{}{ \PrbEsp{;_(;}{Z_j,Z_k}{Z_l}{Z_j\cdot Z_k}{Z_l=z_l}{} }{}{}
  \\\nonumber
  &\equiv&
    \sum\limits_{z_l\in\{-1,0,+1\}}\delim{(}{)}{
      \sum\limits_{(z_j,z_k)\in\{-1,0,+1\}^2}\delim{(}{)}{
        z_j\cdot z_k\cdot \Prb{;(;}{}{}{Z_j=z_j,Z_k=z_k}{Z_l=z_l}{}
      }\cdot\Prb{(}{}{}{Z_l=z_l}{}{}
    }
}

\noindent
that, due to some simplifying hypotheses, we find the previous formula, rewritten in the notation presented

\eqn{}{}{\nonumber
  \PrbEsp{_(}{Z_j,Z_k,Z_l}{}{Z_j\cdot Z_k}{}{}
  =
    \PrbEsp{;_(;}{Z_j,Z_k}{Z_l}{Z_j\cdot Z_k}{Z_l=0}{}\cdot\Prb{_(}{Z_l}{}{0}{}{}
}

Substituting in Bell's inequality, we have

\eqn{0pt}{}{\nonumber
  1-\PrbEsp{;(;}{Z_1,Z_3}{Z_2}{Z_1\cdot Z_3}{Z_2=0}{}\cdot\Prb{_(}{Z_2}{}{0}{}{}
  \geq
    \delim{|}{|}{
      \PrbEsp{;(;}{Z_1,Z_2}{Z_3}{Z_1\cdot Z_2}{Z_3=0}{}\cdot\Prb{_(}{Z_3}{}{0}{}{}
      -\PrbEsp{;(;}{Z_2,Z_3}{Z_1}{Z_2\cdot Z_3}{Z_1=0}{}\cdot\Prb{_(}{Z_1}{}{0}{}{}
    }
}

In the literature we have $\PrbEsp{;(;}{Z_j,Z_k}{Z_l}{Z_j\cdot Z_k}{Z_l=0}{}=\cos(2\cdot(\theta_k-\theta_j))$ therefore $-1\leq\PrbEsp{;(;}{Z_j,Z_k}{Z_l}{Z_j\cdot Z_k}{Z_l=0}{}\leq+1$. Replacing the maximum or minimum values of the values expected conditionals, so as to minimize the first member and maximize the second, we have

\eq{}{\nonumber
  1-\Prb{(}{}{}{Z_2=0}{}{}
  \geq
    \delim{|}{|}{
      \Prb{(}{}{}{Z_3=0}{}{}+\Prb{(}{}{}{Z_1=0}{}{}
    }
}

\noindent
resulting in inequality

\eq{}{\label{eq:LimitesDesigBellProbCond}
  1
  \geq
    \Prb{(}{}{}{Z_2=0}{}{}+\Prb{(}{}{}{Z_3=0}{}{}+\Prb{(}{}{}{Z_1=0}{}{}
}

\noindent
which is certainly satisfied, since each of the probabilities refers to the proportions in which each configuration the experiment is performed, and clearly the sum of all must be 1.

In terms of estimators, we have that the expected values are given by

\eq{}{\label{eq:ProbEstEspCond}
  \PrbEsp{;(;}{j,k}{l}{Z_j\cdot Z_k}{Z_l=0}{}
  \esteq
    \tfrac{
      n_{j,k}^{--}-n_{j,k}^{-+}-n_{j,k}^{+-}+n_{j,k}^{++}
    }{n_{j,k}}
  ,\quad
  n_{j,k}
  =
    n_{j,k}^{--}+n_{j,k}^{-+}+n_{j,k}^{+-}+n_{j,k}^{++}
  \\\label{eq:ProbEstConfig}
  \Prb{(}{}{}{Z_l=0}{}{}
  \esteq
    \tfrac{n_{j,k}}{N}
  ,\quad
  N
  =
    n_{1,2}+n_{1,3}+n_{2,3}
  \\\nonumber
  \PrbEsp{_(}{Z_j,Z_k,Z_l}{}{Z_j\cdot Z_k}{}{}
  \esteq
    \tfrac{n_{j,k}^{--}-n_{j,k}^{-+}-n_{j,k}^{+-}+n_{j,k}^{++}}{N}
}

\noindent
where do we get

\eq{}{\nonumber
  \overbrace{
    \delim{(}{)}{\tfrac{n_{j,k}^{--}-n_{j,k}^{-+}-n_{j,k}^{+-}+n_{j,k}^{++}}{N}}
  }^{\PrbEsp{_(}{Z_j,Z_k,Z_l}{}{Z_j\cdot Z_k}{}{}\esteq}
  =
    \overbrace{
      \delim{(}{)}{\tfrac{n_{j,k}^{--}-n_{j,k}^{-+}-n_{j,k}^{+-}+n_{j,k}^{++}}{n_{j,k}}}
    }^{\PrbEsp{;_(;}{Z_j,Z_k}{Z_l}{Z_j\cdot Z_k}{Z_l=0}{}\esteq}
    \cdot
    \overbrace{
      \delim{(}{)}{\tfrac{n_{j,k}}{N}}
    }^{\Prb{(}{}{}{Z_l=0}{}{}\esteq}
}
\noindent
that substituting Bell's inequality results in inequalities (\ref{eq:BellDesigualdadeEstimadores1}) and, so there is no violation whatever the parameter values.

\end{widetext}

\section{Conclusions}

We conclude, from the estimators (section \ref{Sec:Statistical estimators}) used in the quantum literature and comparing with what is found in the statistical literature, that the expected values used are conditional expected values (section \ref{Sec:Conditional probability}), being necessary to multiply by the probability related to the configuration, thus the inequality is not violated, whatever be the angles of the apparatus. We have also seen that in demonstrating Bell's inequality, Kolmogorov's axioms are used, and that the violation of Bell's and Wigner's inequalities imply a violation of Kolmogorov's axioms.

We observe that Bell and Wigner inequalities are related (\ref{eq:RelacaoBellWigner}) , and the regions (set of angle values) in which there is violation (fig.\ref{fig:DesigBellWignerFisQ}) , each of them coincide. We also note that the attribution of values found in the quantum mechanics literature to marginal probability functions leads us to find negative values (where inequality is violated) for the joint probabilities of $ (Z_1, Z_2, Z_3) $, but inequality, when respecting the conditions used in demonstrating (\ref{eq:DesigBellGeral}) Bell's inequality (ie, when not exchanging $ | \Prb {(} {} {} {\prbs} { } {} | $ for $ \Prb {(} {} {} {\prbs} {} {} $), there is no violation of Bell's inequality.

We show in section \ref{Loopholes} that it is possible to model more general cases, being able to assign probabilities for cases in which there is only one detection or none, and even the case in which there are 3 detections, thus covering cases in which detectors do not have 100$ \% $ efficiency and cases in which there are interactions between the devices and the external environment. We saw that, regardless of the efficiency of the apparatus and the isolation with respect to the external environment, whatever the variables that occurred in the values of the probabilities, in this modeling, Bell's inequality was never violated (\ref{eq:LimitesDesigBellProbCond}). Such modeling (\ref{eq:ProbCond})(\ref{eq:Valor esperado e valor esperado condicional}), besides being broader, consistent with the Probability Theory and respects all the conditions used in demonstrating Bell's inequality.

Therefore, in this article, we present the demonstration of Bell's inequality found in the literature, to later present our studies on the conditions used in the articles and also the probability axioms used, and thus we clarify which conditions (and axioms) are violated, specifically the relationship between estimators used with conditional probabilities, we extend these conditions to cover other situations and thus, in this way, we show that Bell's inequality is compatible with the Probability Theory.

\end{document}